# TD-CARMA: Painless, Accurate, and Scalable Estimates of Gravitational Lens Time Delays with Flexible CARMA Processes

Antoine D. Meyer[1,2], David A. van Dyk[1], Hyungsuk Tak[3,4,5], and Aneta Siemiginowska[2]
[1] Statistics Section, Department of Mathematics, Imperial College London, 180 Queen's Gate, London SW7 2AZ, UK; a.meyer19@imperial.ac.uk
[2] Center for Astrophysics | Harvard & Smithsonian, 60 Garden Street, Cambridge, MA 02138, USA
[3] Department of Statistics, Pennsylvania State University, University Park, PA 16802, USA
[4] Department of Astronomy and Astrophysics, Pennsylvania State University, University Park, PA 16802, USA
[5] Institute for Computational and Data Sciences, Pennsylvania State University, University Park, PA 16802, USA

## Abstract

Cosmological parameters encoding our understanding of the expansion history of the universe can be constrained by the accurate estimation of time delays arising in gravitationally lensed systems. We propose TD-CARMA, a Bayesian method to estimate cosmological time delays by modeling observed and irregularly sampled light curves as realizations of a continuous auto-regressive moving average (CARMA) process. Our model accounts for heteroskedastic measurement errors and microlensing, an additional source of independent extrinsic long-term variability in the source brightness. The semiseparable structure of the CARMA covariance matrix allows for fast and scalable likelihood computation using Gaussian process modeling. We obtain a sample from the joint posterior distribution of the model parameters using a nested sampling approach. This allows for "painless" Bayesian computation, dealing with the expected multimodality of the posterior distribution in a straightforward manner and not requiring the specification of starting values or an initial guess for the time delay, unlike existing methods. In addition, the proposed sampling procedure automatically evaluates the Bayesian evidence, allowing us to perform principled Bayesian model selection. TD-CARMA is parsimonious, and typically includes no more than a dozen unknown parameters. We apply TD-CARMA to six doubly lensed quasars HS2209+1914, SDSS J1001+5027, SDSS J1206+4332, SDSS J1515+1511, SDSS J1455+1447, and SDSS J1349+1227, estimating their time delays as $-21.96 \pm 1.448$, $120.93 \pm 1.015$, $111.51 \pm 1.452$, $210.80 \pm 2.18$, $45.36 \pm 1.93$, and $432.05 \pm 1.950$, respectively. These estimates are consistent with those derived in the relevant literature, but are typically two to four times more precise.

*Unified Astronomy Thesaurus concepts:* Astrostatistics techniques (1886); Astrostatistics tools (1887); Bayesian statistics (1900); Quasars (1319); Gravitational lensing (670); Nested sampling (1894); Light curves (918); Time series analysis (1916)

## 1. Introduction

Cosmological parameters describing the evolution and current state of the universe through the Λ-cold dark matter (ΛCDM) model, such as the Hubble constant $H_0$, are measured by combining baryonic acoustic oscillations measurements and cosmic microwave background observations (Ade et al. 2014). However, cosmological probes are subject to statistical and systematic errors, and individual probes are not able to constrain the cosmological models uniquely (Rathna Kumar et al. 2013). Combining parameter constraints obtained from different techniques can help break degeneracies associated with specific methods (Tewes et al. 2013). In particular, Refsdal (1964) shows that independent constraints on the cosmological parameters can be derived from the measurement (or estimation) of time delays in quasars subject to strong gravitational lensing by a foreground galaxy. This time delay cosmography (Treu & Marshall 2016; Treu et al. 2022; Birrer et al. 2022) has since become an important cosmological probe, especially in the presence of the Hubble tension (Verde et al. 2019; Shah et al. 2021).

Observations of lensed quasar systems are increasingly available, and the upcoming Vera C. Rubin Observatory is expected to monitor around 1000 strongly lensed quasars (Abell et al. 2009; Dobler et al. 2015). This has prompted the development of numerical and statistical analysis tools for cosmological time delays, notably encouraged by the 2015 strong lens time delay challenge (TDC; Dobler et al. 2015; Liao et al. 2015). The accurate estimation of cosmological time delays has been challenging because of several unusual features in light-curve data. First, celestial cycles and observational patterns cause light-curve data to be irregularly sampled and to feature large seasonal gaps. Second, the observed magnitudes of light curves are subject to heteroskedastic measurement errors. Finally, brightness fluctuations in the observed light curves are induced by multiple sources of variability, which all should be taken into account for meaningful modeling and accurate estimation. Indeed, in addition to the intrinsic variability in the source, strong gravitational lensing magnifies the brightness of the light curves, and a microlensing effect may independently encode extrinsic long-term variability in the observed light curves.

Cosmological time delay estimation has been an active research area in astronomy for decades. We refer readers to Linder (2011), Treu & Marshall (2016), Suyu et al. (2017), and Treu et al. (2022) for an overview of time delay cosmography, and to Eulaers et al. (2013), Tewes et al. (2013), Liao et al.

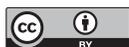

(2015), and Tak et al. (2017) for detailed methodological presentations. See also Courbin et al. (2018), Bonvin et al. (2016, 2018, 2019), and Millon et al. (2020a) for recent advances in time delay estimation methodology.

In a broad sense, time delay estimation techniques can be categorized into nonparametric and parametric methods. Nonparametric time delay estimation methods minimize a measure of difference between the lensed light curves directly from the data and provide uncertainty estimates using Monte Carlo simulations (Pelt et al. 1994). Parametric techniques, on the other hand, model the stochastic variability of active galactic nucleus (AGN) light curves within either a Bayesian or a likelihood-based framework.

In the nonparametric family of estimation techniques, Aghamousa & Shafieloo (2015) maximize the cross-correlation between smoothed versions of the lensed light curves to estimate the time delay, and produce error estimates through simulations. Similarly, Rathna Kumar et al. (2013) minimize the residuals of a high-pass filtered difference light curve between the lensed quasar light curves. Bag et al. (2022) propose a Gaussian-smoothing based nonparametric method to identify and measure time delays in lensed quasar images. In the parametric family of techniques, examples include Hojjati et al. (2013) who perform Gaussian process (GP) regression to model the intrinsic light curve and Millon et al. (2020b) who use a free-knot spline technique. More recently, Bag et al. (2021) developed a Bayesian method to identify gravitationally lensed Type Ia supernovae (SNe Ia) from unresolved light curves and measure their time delays, and Donnan et al. (2021) propose a time delay estimation method based on a running optimal average (ROA) model for the light curve data.

In this paper, we build upon and address limitations of existing Bayesian techniques, using the method developed in Tak et al. (2017) as our main reference. The Bayesian inference paradigm allows us to quantify directly the uncertainty in the time delay parameter. Tak et al. (2017) develop a Bayesian method and apply it to estimate time delays in doubly and quadruply lensed quasar light curves. They model the intrinsic quasar light curve as a realization of a damped random walk (DRW) process (usually referred to as the Ornstein–Uhlenbeck process in the statistical literature) and assume that the light curves are time- and magnitude-shifted copies of one another. Hereafter we label this method as TD-DRW. Microlensing is also directly accounted for in TD-DRW, and is modeled as a polynomial regression, usually of third order (Kochanek et al. 2006; Courbin et al. 2011; Morgan et al. 2012; Tak et al. 2017). The reader is referred to Tewes et al. (2013) for a review on techniques designed to handle extrinsic variability due to microlensing in time delay estimation problems.

The TD-DRW method performed well in the TDC (Liao et al. 2015), perhaps in part because the TDC data had been generated under a DRW process. The aim of this paper is to develop this method further in order to improve the accuracy of time delay estimation, by refining the intrinsic quasar light-curve model and addressing several additional limitations.

Following the introduction of the DRW process as a modeling tool for the stochastic variability in AGN light curves by Kelly et al. (2009), numerous empirical studies have supported its use in astronomical time series data analysis (MacLeod et al. 2010; Kozłowski et al. 2009; Kim et al. 2012; Andrae et al. 2013). However, the inflexibility of DRW (which only incorporates two free parameters) has limited its broader use (Mushotzky et al. 2011; Zu et al. 2013; Graham et al. 2014; Kasliwal et al. 2015; Kozłowski 2016, 2017), and has motivated the adoption of more general models. Specifically, Kelly et al. (2014) introduced the continuous auto-regressive moving average (CARMA) processes as a flexible tool for characterizing AGN variability. Recent astrophysical studies (Kasliwal et al. 2017; Moreno et al. 2019; Ryan et al. 2019; Stone et al. 2022) have empirically shown that the CARMA process can successfully model various types of stochastic variability in AGN light curves.

Indeed, the power spectrum density (PSD) of a CARMA process can account for multiple break-like features and quasiperiodic oscillations (QPOs) in their PSDs, which are characteristics typically observed in AGN data (Kelly et al. 2014; Kasliwal et al. 2017; Ryan et al. 2019). As explained in Liao et al. (2015), AGNs and particularly quasars are sources for which strong gravitational lensing is more likely to happen, since they are very distant from Earth (thus increasing the likelihood that a candidate lens stands between the source and Earth) and very luminous. Therefore, we update the TD-DRW method introduced in Tak et al. (2017) by generalizing the intrinsic light-curve model from a DRW process to CARMA processes, and hereafter refer to our method as TD-CARMA.

In addition to leveraging CARMA processes to perform a finer modeling of the intrinsic light curve, we address computational limitations of existing time delay estimation methods. For example, Kelly et al. (2014) report that the likelihood function (and thus the resulting joint posterior distribution) of the CARMA model may contain multiple local maxima, especially when the degree of the auto-regression component is greater than one. It is also well known that time delay estimates under the TD-DRW may suffer from multi-modality (Tak et al. 2017, 2018). Because it combines two sources of multimodality, namely a CARMA model and time delay, it is not unexpected that TD-CARMA exhibits more severe multimodality in its posterior distribution.

To address this, we perform Bayesian parameter inference using nested sampling (Skilling 2006) instead of the Markov Chain Monte Carlo (MCMC) methods used by Kelly et al. (2014) and Tak et al. (2018). Specifically, we use the pyMultiNest (Buchner et al. 2014) python implementation of the multimodal ellipsoidal nested sampling algorithm developed by Feroz et al. (2009), known as MultiNest.

MultiNest has two important computational advantages over MCMC in this setting. First, most time delay estimation methods require and are highly sensitive to an initial guess for the time delay parameter. For instance, in TD-DRW, Tak et al. (2017) compute the prohibitively expensive profile likelihood of the time delay parameter to find a starting value for their MCMC algorithm, and all methods described in Tewes et al. (2013) must be initialized near a plausible value of the time delay. Unlike MCMC, however, nested sampling methods do not evolve a single path through the parameter space and thus do not require the input of an initial value for the model parameters. This means TD-CARMA only requires the specification of a range of possible values for each of the parameters. Also, MultiNest computes the Bayesian evidence of the fitted model simultaneously to the posterior sampling, which solves the model selection problem of choosing the auto-regressive and moving average orders of the CARMA process, as well as the degree of the polynomial regression used to model microlensing, without any further computation.

The improvements that TD-CARMA offers over existing methods introduced in Tewes et al. (2013), Liao et al. (2015), and Tak et al. (2017) can be summarized as follows.

1. *Flexibility and accuracy.* TD-CARMA leverages flexible CARMA processes which are well suited to model a wide range of AGN light curves. In addition, this work shows that these flexible models can also improve the accuracy of time delay estimates.
2. *Painless Bayesian computation.* Using MultiNest for Bayesian computation means that TD-CARMA does not require a plausible initial value for the time delay parameter. MultiNest also allows us to tackle the severe multimodality in the joint posterior distribution of the TD-CARMA model parameters.
3. *Scalability*. The model likelihood is efficiently computed with linear complexity $O(n)$ using GP modeling. This makes TD-CARMA scalable to large data sets.
4. *Built-in model selection.* The proposed MultiNest fit automatically evaluates the Bayesian evidence of the fitted models and allows us to perform principled Bayesian model comparison and selection without any additional computation.
5. *Parsimony.* Some methods featured in Liao et al. (2015) and Tewes et al. (2013) incorporate hundreds of parameters, and sometimes have parameter spaces that scale with the data. In contrast, in our numerical studies the number of unknown model parameters used by TD-CARMA ranges from 6 to 15. Parsimonious methods are easier to interpret scientifically and also statistically, since better behaved likelihoods produce more Gaussian estimators, and more easily quantifiable errors.

The rest of the paper is organized as follows. Section 2 presents the time delay estimation method TD-CARMA. Section 3 details our Bayesian inferential procedure, including MultiNest. In Section 4, we show through simulation studies that the CARMA-based time delay estimation method is more applicable and accurate than several popular methods. In Section 5, we apply the method to doubly lensed quasars HS2209+1914, Sloan Digital Sky Survey (SDSS) J1001 +5027, SDSS J1206+4332, SDSS J1515+J1511, SDSS J1455+1447, and SDSS J1349+1227. Finally, Section 6 presents further research directions and concludes. Background material and detailed results from our numerical studies appear in a number of appendices.

## 2. A CARMA-based Time Delay Model

Strong gravitational lensing can produce multiple copies of the observed source's light curve. While our statistical model is explicitly designed for the doubly lensed case, it can be applied pairwise to more complex systems (as in Tak et al. 2017) and could be generalized to the quadruply lensed case in order to account for correlations among the time delay parameters. Henceforth, however, we consider the doubly lensed case.

### 2.1. Data

The data for a doubly lensed quasar observed for $n$ epochs can be written as $\boldsymbol{D} = \{t_i, x_i, \delta_i^x, y_i, \delta_i^y\}_{i=1}^n$, where $t_i$ denotes the measurement time (in days) of observation epoch $i$, $x_i$ and $y_i$ denote the observed brightness of the light curves of the two lensed images of the same quasar at time $t_i$ (typically measured in apparent magnitude), and $\delta_i^x$ and $\delta_i^y$ respectively denote the standard deviation of the measurement error on $x_i$ and $y_i$ (thus encoding the heteroskedasticity of measurement errors). Table 1 summarizes the symbols used in this work. Figure 1 shows the data for the doubly lensed quasar SDSS J1001 +5027, where light curve B is delayed from light curve A, for $\Delta \approx 119.3$ days according to previous analyses (a figure reproduced from Kumar et al. 2013).

**Table 1**
Glossary of the Notation Used in This Work

| Data | |
|---|---|
| Symbol | Description |
| $t$ | Observation times |
| $x$ | Brightness measurements for light curve A |
| $y$ | Brightness measurements for light curve B |
| $z$ | Composite light curve |
| $\delta^x$ | Standard deviations for light curve A |
| $\delta^y$ | Standard deviations for light curve B |
| **Model Parameters** | |
| Symbol | Description |
| $\Delta$ | Time delay (days) |
| $\theta$ | Microlensing polynomial regression coefficients |
| $\alpha$ | CARMA auto-regressive coefficients |
| $\beta$ | CARMA moving average coefficients |
| $\tau$ | DRW mean-reversion timescale |
| $\sigma^2$ | CARMA and DRW white noise variance |
| $\mu$ | CARMA and DRW long-term mean |
| $\Omega$ | Full set of either DRW or CARMA parameters |
| **Model Hyperparameters**[a] | |
| Symbol | Description |
| $p$ | Auto-regressive order |
| $q$ | Moving average order |
| $m$ | Microlensing polynomial regression order |
| **Other Symbols** | |
| Symbol | Description |
| $\mathcal{Z}$ | Bayesian evidence |
| $\mathcal{Z}_i$ | Bayesian evidence for Model $i$ or Mode $i$ |
| $\pi_i$ | Relative probability of Model $i$ or Mode $i$ |
| $\phi$ | Generic parameter vector |
| $\Phi$ | Range of possible values for generic parameter $\phi$ |

**Note.**
[a] The hyperparameters index competing models. Values for the hyperparameters are selected via the Bayesian evidence or the relative probabilities of the models considered; see Section 3.1.

### 2.2. Time Delay Estimation Framework

The general idea of our time delay estimation framework is to reconstruct the common intrinsic quasar light curve from the measurements of its two lensed counterparts, under the assumption that they are time- and magnitude-shifted versions of one another, up to a microlensing term, where the time shift corresponds to the time delay $\Delta$.

The magnitude measurements $\boldsymbol{x} = \{x_i\}_{i=1}^n$ and $\boldsymbol{y} = \{y_i\}_{i=1}^n$ are assumed to be discrete realizations of unobserved continuous light curves $X(t)$ and $Y(t)$, which represent the true source magnitudes at time $t \in \mathbb{R}$. As a result of the strong gravitational lensing phenomenon, we assume that $Y(t)$ is a time-shifted version of $X(t)$, which yields

$$Y(t) = X(t - \Delta), \quad (1)$$

where $\Delta$ is the time delay in days. In addition to the intrinsic brightness fluctuations of the source, two lensing effects induce extra variability in the observed magnitudes. First, strong gravitational lensing magnifies the brightness of each light curve to different degrees. Therefore $x$ and $y$ can exhibit different average magnitudes. Tak et al. (2017) take this effect into account by introducing an intercept term $\theta_0$ as follows

$$Y(t) = X(t - \Delta) + \theta_0. \quad (2)$$

Second, additional extrinsic long-term variability might be caused by microlensing, a lensing effect occurring when a light ray passes close to a moving object near the lensing galaxy. This effect is typically modeled by a polynomial regression of order $m$ on time (Kochanek et al. 2006; Courbin et al. 2011; Morgan et al. 2012; Tak et al. 2017). That is

$$Y(t) = X(t - \Delta) + \boldsymbol{w}_m(t - \Delta)\boldsymbol{\theta}, \quad (3)$$

where $\boldsymbol{w}_m(t - \Delta) = \{1, t - \Delta, ..., (t - \Delta)^m\}$ is the vector containing the polynomial time variables and $\boldsymbol{\theta} = \{\theta_0, ..., \theta_m\}$ is the microlensing (regression) coefficients. We treat $m$ as a hyperparameter. That is, it indexes the choice of model and its value is set by comparing several values as part of the statistical model selection, see Section 3.1.

Given the parameters $\Delta$ and $\boldsymbol{\theta}$, we can construct a composite light curve that combines two lensed light curves (as if they were a single light curve) by adjusting one light curve according to the model in Equation (3). The combined observation times are composed of the original observation times and the time-delay-shifted observation times, $\boldsymbol{t}^\Delta = \{t_i\}_{i=1}^n \cup \{t_i - \Delta\}_{i=1}^n$. We denote the magnitudes of the composite light curve by $\boldsymbol{z} = \{z_j\}_{j=1}^{2n}$, where each $z_j$ is defined as

$$z_j = \begin{cases} x_i & \text{for some } i \text{ if } t_j^\Delta \text{ is in } \boldsymbol{t}, \\ y_i - \boldsymbol{w}_m(t_j - \Delta)\boldsymbol{\theta} & \text{for some } i \text{ if } t_j^\Delta \text{ is in } \boldsymbol{t} - \Delta. \end{cases} \quad (4)$$

We can similarly define the vector of measurement error standard deviations as $\boldsymbol{\delta}^z = \{\delta_i^z\}_{i=1}^{2n}$ with

$$\delta_j^z = \begin{cases} \delta_i^x & \text{for some } i \text{ if } t_j^\Delta \text{ is in } \boldsymbol{t}, \\ \delta_i^y & \text{for some } i \text{ if } t_j^\Delta \text{ is in } \boldsymbol{t} - \Delta. \end{cases} \quad (5)$$

We assume that the vector of measurements for the composite light curve $\boldsymbol{z}$ is the discrete realization of an unobserved continuous-time light curve $Z(t)$, which represents the true source magnitude of the AGN at time $t \in \mathbb{R}$. Therefore, given $\Delta$ and $\boldsymbol{\theta}$, we only need to model the stochastic variability of the single latent light curve $Z(t)$. The DRW and CARMA processes have been widely used to model the stochastic nature of the intrinsic variability in AGN light curves since Kelly et al. (2009, 2014) introduced them to the astronomical community.

### 2.3. DRW and CARMA Processes

The DRW process has already been integrated into time delay estimation methodology by Tak et al. (2017), and we propose an extension of this method by modeling the intrinsic light curve $Z(t)$ with the more general CARMA processes.

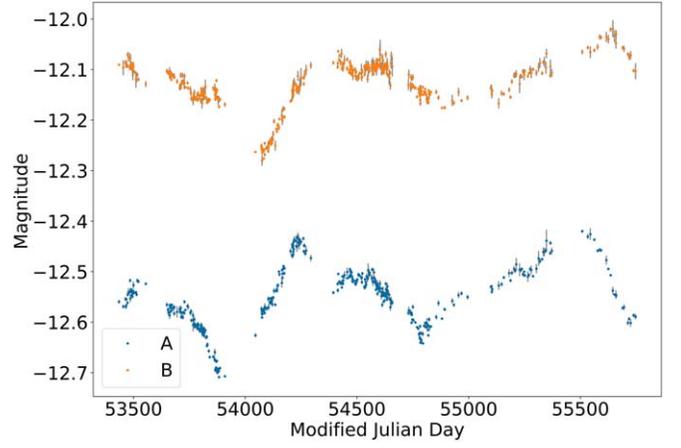

**Figure 1.** Data for the doubly lensed quasar SDSS J1001+5027 Kumar et al. (2013), studied in Section 5.3 of this work.

The DRW process is defined as the solution of the stochastic differential equation

$$dZ(t) = -\frac{1}{\tau}(Z(t) - \mu)dt + \epsilon(t), \quad (6)$$

where $\mu$ is the long-term mean of the process, $\tau$ is the timescale for the process to revert to its long-term mean, and $\epsilon(t)$ is a white noise process with variance $\sigma^2$ (that is, $\epsilon(t) \sim N(0, \sigma^2)$) governing the stochastic fluctuations in the process (Brockwell & Davis 2002).

A CARMA process is indexed by two hyperparameters: its auto-regressive order $p$ and moving average order $q$. The process is denoted CARMA($p$, $q$) and is defined as the solution of the stochastic differential equation

$$\frac{d^p Z(t)}{dt^p} + \alpha_{p-1}\frac{d^{p-1} Z(t)}{dt^{p-1}} + ... + \alpha_0 Z(t)$$
$$= \mu dt + \beta_q \frac{d^q \epsilon(t)}{dt^q} + \beta_{q-1}\frac{d^{q-1} \epsilon(t)}{dt^{q-1}} + ... + \epsilon(t), \quad (7)$$

where $\boldsymbol{\alpha} = \{\alpha_0, ..., \alpha_{p-1}\}$ are the auto-regressive parameters, $\boldsymbol{\beta} = \{\beta_1, ..., \beta_q\}$ are the moving average coefficients, and $\epsilon(t)$ is a white noise process with variance $\sigma^2$. We also define $\alpha_p = 1$ and $\beta_0 = 1$, and note that the DRW is equivalent to the CARMA(1,0) process with $\tau = 1/\alpha_0$.

DRW has a limited applicability in describing the stochastic variability in AGN light curves, since for instance it cannot account for quasiperiodic variability features and multiple auto-correlation timescales. CARMA processes, however, can account for such features and are thus suitable to characterize the variability in a wide range of AGN light curves. The reader is referred to Appendix A, Kelly et al. (2009), and Kelly et al. (2014) for technical reviews of the mathematics governing the dynamics of the DRW and CARMA processes.

### 2.4. Time Delay Model Likelihood

DRW and CARMA processes that are stochastically driven by white noise processes are Gaussian (Kelly et al. 2009, 2014)[6] and their likelihood functions can be evaluated via a GP. Let $\Omega$ be the parameter vector for the intrinsic light curve $Z(t)$, i.e., $\Omega = (\mu, \sigma^2, \tau)$ for DRW and $\Omega = (\mu, \sigma^2, \boldsymbol{\alpha}, \boldsymbol{\beta})$ for CARMA. GPs are formulated in terms of a mean function $M(t_i)$ (parametrized by $\mu$, $\boldsymbol{\theta}$, and $\Delta$ in our case), and a covariance or kernel function $k(t_i, t_j)$ (parametrized by $\sigma^2$ and $\tau$ for DRW and by $\boldsymbol{\alpha}$ and $\boldsymbol{\beta}$ for CARMA). Following the notation in Foreman-Mackey et al. (2017), we can write the log-likelihood function of the model parameters given observed data as

$$\ln L(\Delta, \boldsymbol{\theta}, \Omega | \mathbf{z}) = \ln p(\mathbf{z} | \Delta, \boldsymbol{\theta}, \Omega)$$
$$= -\frac{1}{2}\mathbf{r}^T K^{-1} \mathbf{r} - \frac{1}{2}\ln\det(K) - N\ln(2\pi), \quad (8)$$

where

$$\mathbf{r} = (z_1 - M(t_1^\Delta), ..., z_{2n} - M(t_{2n}^\Delta))^T, \quad (9)$$

and the elements of the covariance matrix $K$ are $[K]_{ij} = k(t_i^\Delta, t_j^\Delta) = R(t_i^\Delta - t_j^\Delta)$, with $R(\cdot)$ defined in Equation (A5) of Appendix A.

---

[6] CARMA processes can be defined as stochastically driven by non-Gaussian Levy processes, in which case they are not Gaussian (Brockwell 2001).

The evaluation of a GP likelihood function given a realization of $n$ data points requires the inversion of matrix $K$, which computationally scales as $O(n^3)$. This represents a prohibitive bottleneck that limits the computational scalability to large data sets. However, Foreman-Mackey et al. (2017) developed an algorithm exploiting the semiseparable structure of CARMA covariance matrices, enabling fast computation of the likelihood function with only linear complexity $O(n)$. The likelihood function of a CARMA process may also be computed using the Kalman filter algorithm (Jones 1981; Jones & Ackerson 1990; Kelly et al. 2014), and although both methods have the same computational time complexity, the GP-based algorithm has been found to be an order of magnitude faster in practice (Foreman-Mackey et al. 2017). We use the EzTao package (Yu & Richards 2022; Yu et al. 2022) for the GP-based computation of our model likelihood. A python package, TD-CARMA, is publicly available to implement our proposed method.

The reader is referred to Williams & Rasmussen (2006), Foreman-Mackey et al. (2017), and Yu et al. (2022) for reviews on the GPs and GP-based computations for CARMA processes.

## 3. Bayesian Inference and Statistical Computation

### 3.1. Bayesian Inference and Model Selection

We adopt a Bayesian statistical approach to perform parameter inference, which allows us to quantify the uncertainties in the model parameters, including $\Delta$, via their joint posterior distribution, given the observed data. From this we can compute point estimates and error bars, as required. Once the prior distributions are specified and the data are observed (and incorporated into the model via the likelihood function), Bayes' Theorem gives the expression for the posterior distribution of the parameters, i.e., the updated summary of information about the model parameters after observing the data

$$p(\Delta, \boldsymbol{\theta}, \Omega | \boldsymbol{D}) = \frac{L(\Delta, \boldsymbol{\theta}, \Omega | \boldsymbol{D}) p(\Delta, \boldsymbol{\theta}, \Omega)}{p(\boldsymbol{D})}, \quad (10)$$

where $L(\Delta, \boldsymbol{\theta}, \Omega | \boldsymbol{D})$ is the likelihood under the model defined in Equation (8) and is equal to $p(\boldsymbol{D} | \Delta, \boldsymbol{\theta}, \Omega)$, the sampling distribution of the data given the model parameters; $p(\Delta, \boldsymbol{\theta}, \Omega)$ is the prior distribution of the parameters, and $p(\boldsymbol{D})$ is the marginal distribution of the data, otherwise known as the Bayesian evidence. The Bayesian evidence, hereafter denoted by $\mathcal{Z}$, i.e., $\mathcal{Z} = p(\boldsymbol{D})$, can be used for model comparisons, and is defined as

$$\mathcal{Z} = \int_\Phi p(\boldsymbol{D}|\phi) p(\phi) d\phi, \quad (11)$$

where $\phi$ is a generic (vector) parameter and $\Phi$ is the (multivariate) set of possible values for $\phi$. The Bayesian evidence $\mathcal{Z}$ is a measure of the "goodness of fit" of a model to the data $\boldsymbol{D}$, so that models with higher evidence provide a better explanation of the data. The definition of $\mathcal{Z}$ naturally incorporates Occam's razor, as more complex models, i.e., models defined on higher-dimensional parameter spaces, are penalized if they do not sufficiently improve the fit to the data (Jefferys & Berger 1991).

Suppose we fit each of several models, $M_1, ..., M_K$ and wish to identify the models that dominate in terms of their evidence. For example, we may wish to compare TD-DRW($m$) and TD-CARMA($p$, $q$, $m$) for each of several combinations of values of the hyperparameters, $p$, $q$, and $m$ (auto-regressive, moving average, and microlensing orders).

For this, it is useful to compute the relative probability of each considered model

$$p(M_i) = \frac{\mathcal{Z}_i}{\sum_{k=1}^K \mathcal{Z}_k}. \quad (12)$$

In TD-CARMA we must specify the hyperparameter triplet ($p$, $q$, $m$) that best fits the data. Statistically, this is a model selection problem. Because CARMA($p$, $q$) models are nonnested, likelihood ratio tests cannot be used for model selection (Protassov et al. 2002; Kelly et al. 2014). In the astrophysics literature, when fitting CARMA processes to light-curve data, ($p$, $q$) pairs are selected using information criteria such as the Akaike information criterion (AIC; Kelly et al. 2014) or deviance information criterion (DIC; Kasliwal et al. 2017). These may require additional computation (e.g., AIC requires the computation of the maximum likelihood estimator) and rely on asymptotic approximations. We instead use the principled Bayesian evidence, $\mathcal{Z}$, which is a commonly used model selection approach in the Bayesian regime (mostly via Bayes' factors; e.g., Kass & Raftery 1995). The Bayesian evidence is conveniently computed directly by MultiNest[7] at the same time it generates a posterior sample, see Section 3.3.

### 3.2. Prior Distributions

We assume uniform prior distributions for each model parameter. Specifically, we set a uniform prior for the time delay $\Delta$ in the range $[t_1 - t_n, t_n - t_1]$. In instances where it is clear which light curve is delayed, the prior distribution for $\Delta$ can accordingly be reduced to $[t_1 - t_n, 0]$ or $[0, t_n - t_1]$. The prior for the microlensing parameters, $\boldsymbol{\theta}$, is defined to be uniform on the range $[-10^5, 10^5]$. The average apparent magnitude $\mu$ is given a uniform prior on the range $[-30, 30]$, which covers the magnitude range from that of the Sun to that of the faintest object observable by the Hubble Space Telescope (Tak et al. 2017). The other TD-CARMA parameters $\boldsymbol{\alpha}$, $\boldsymbol{\beta}$, and $\sigma$ are sampled on the logarithmic scale, with priors on their (natural) logs defined to be uniform on the range [−15, 15]. This largely covers the range of frequencies probed by the data.[8]

---

[7] MultiNest returns the logarithm of the Bayesian evidence, i.e., $\ln(\mathcal{Z}_i)$.

### 3.3. Multiple Modes and MultiNest

We obtain a sample from the posterior distribution $p(\Delta, \beta, \Omega|D)$ in Equation (10) via the MultiNest implementation of (multimodal ellipsoidal) nested sampling (Feroz et al. 2009; Buchner et al. 2014), for both TD-DRW and TD-CARMA. In the case of time delay estimation, inference via nested sampling has a variety of advantages over traditional MCMC methods, including the computation of the Bayesian evidence to perform model selection (see Section 3.1), and the efficient sampling of multimodal posterior distributions.

The multimodality in this problem is complicated, because it manifests itself in two distinct ways.

First, the posterior distribution of CARMA($p$, $q$) parameters is notoriously multimodal when $p > 1$ (Kelly et al. 2014). Moreover, even for fixed $p$ and $q$ there is potential multimodality in the time delay parameters $\Delta$ and $\theta$ (e.g., Tak et al. 2017).

The sampling procedures in Tak et al. (2017) have difficulties navigating such posteriors even with the simpler TD-DRW model. Thus, we use MultiNest, which is specifically designed to sample efficiently from complicated and multimodal posterior distributions. MultiNest identifies the multiple modes in the posterior, partitions the parameter space into regions in which the separated modes are supported, and evaluates the local evidence of each identified mode.

Specifically, assuming MultiNest identifies mode $i$ defined in region $\Phi_i$ of the parameter space $\Phi$, the local evidence $\mathcal{Z}_i$ of mode $i$ is defined as

$$\mathcal{Z}_i = \int_{\Phi_i} p(D|\phi)p(\phi)d\phi. \quad (13)$$

The local evidence $\mathcal{Z}_i$ can be used to compute the relative probability of mode $i$, denoted as $\pi_i$ and defined as

$$\pi_i = \int_{\Phi_i} p(\phi|D)d\phi = \frac{1}{\mathcal{Z}} \int_{\Phi_i} p(D|\phi)p(\phi)d\phi = \frac{\mathcal{Z}_i}{\mathcal{Z}},$$

where $\mathcal{Z}$ is the (total) Bayesian evidence.

In the event of a multimodal posterior distribution, computing the above quantities offers a principled way of quantifying the relative probability of the several modes.

### 3.4. Constraints and Parameterizations

As discussed in Section A.2, we use an alternative parameterization of the auto-regressive polynomial $A(z)$ to enforce stationarity when fitting CARMA processes. Instead of sampling the auto-regressive parameters $\alpha$, we sample the coefficients $a = \{a_1,..., a_p\}$ (defined in Section A.2), on the natural log-scale to enforce nonnegativity (for stationarity of the resulting CARMA process).

The likelihood of a CARMA process is invariant to permutations in the indices of the roots of the auto-regressive polynomial. In order to avoid identifiability issues, Kelly et al. (2014) imposes ordering constraints on the auto-regressive roots. This, however, involves rejecting Monte Carlo draws that do not satisfy the ordering constraint, which renders sampling inefficient. Since swaps in the indices of the auto-regressive roots that give the same likelihood create mirror modes in the posterior distribution, we do not impose an ordering constraint but rather combine the modes and consider the full posterior in our analyses.

Finally, the polynomial regression coefficients $\theta = \{\theta_1,..., \theta_m\}$ are highly correlated in their posterior distributions, since the regression covariates $\{1, t - \Delta, (t - \Delta)^2,..., (t - \Delta)^m\}$ are themselves highly correlated. This yields a posterior distribution with complex geometry that is difficult to sample from efficiently. To circumvent this issue, we perform a QR decomposition[9] (e.g., Gander 1980) of the polynomial regression covariate matrix, a common computational trick in standard regression problems used to decorrelate covariates.

At each iteration, MultiNest proposes new values of the parameters by sampling from their joint prior distribution, constrained to regions of the parameter space corresponding to isocontours in the likelihood (Feroz et al. 2009). The resolution at which the parameter space is explored therefore depends (at least initially) on the number of points sampled from the prior at each iteration, denoted $N_{\text{live}}$ (Feroz et al. 2009), and the range covered by the prior. We repeatedly fitted our models with different values of $N_{\text{live}}$, and found that $N_{\text{live}} = 400$ for our simulations and $N_{\text{live}} = 1000$ for our data set applications yield stable results.

### 3.5. Frequencies in the Fit

In the event of a multimodal posterior distribution of the CARMA($p$, $q$) model parameters, $\Omega = \{\alpha, \beta, \sigma\}$, one or more of the modes might correspond to a quasiperiodic frequency (QPO) in the posterior PSD (i.e., the PSD features a Lorentzian centered away from 0). For example, the left panel of Figure 2 shows a bimodal PSD, where mode 1 detects a frequency at ≈2 days. This posterior PSD was obtained by fitting TD-CARMA (3, 1, 3) to the doubly lensed quasar data set SDSS J1001+5027, the full analysis of which appears in Section 5.3.

Following Kelly et al. (2014), we can attribute a detected frequency in the inferred PSD of the intrinsic light curve to either the signal or the noise by comparing the posterior distribution of the power for the detected frequency with the measurement noise level.[10] If the posterior distribution of power for the detected frequency is supported above the measurement noise level, we recognize the frequency as a feature of the light curve. Otherwise, this provides statistical evidence that the CARMA processes are essentially identifying a periodic artifact of the noise.

If a detected frequency were to be attributed to the intrinsic light curve, it would dramatically reduce the size of the error bars for the estimate of $\Delta$, as shown in the right panel of Figure 2 (and Tables 12, 14, and 15 in Appendix C). Indeed, when a frequency is detected in the PSD for the intrinsic light curve, this naturally constrains $\Delta$ to multiples of the frequency. Thus, TD-CARMA can leverage a frequency that is not attributed to the noise to reduce the error bars on $\Delta$ dramatically. This would not be possible with TD-DRW, for instance.

---

[8] The maximum and minimum frequencies probed by the data are defined as $1/\min(\Delta t)$ and $1/(\max(t) - \min(t))$, respectively, where $\Delta t := \{t_{i+1} - t_i\}_{i=1}^{2n-1}$.

[9] The name "QR decomposition" refers to the notation used in the standard representation of the decomposition, where $Q$ is an orthogonal matrix and $R$ is an upper triangular matrix. We refer readers to the Stan User's Guide, https://mc-stan.org/docs/2_29/stan-users-guide/QR-reparameterization.html, for a complete description of the QR decomposition and its use in decorrelating coefficients in linear statistical models.

[10] The measurement noise level is computed as $2\bar{\Delta}_t \bar{\delta}^2$, where $\bar{\Delta}_t$ is the average time difference between consecutive time measurements and $\bar{\delta}^2$ is the average measurement error variance.

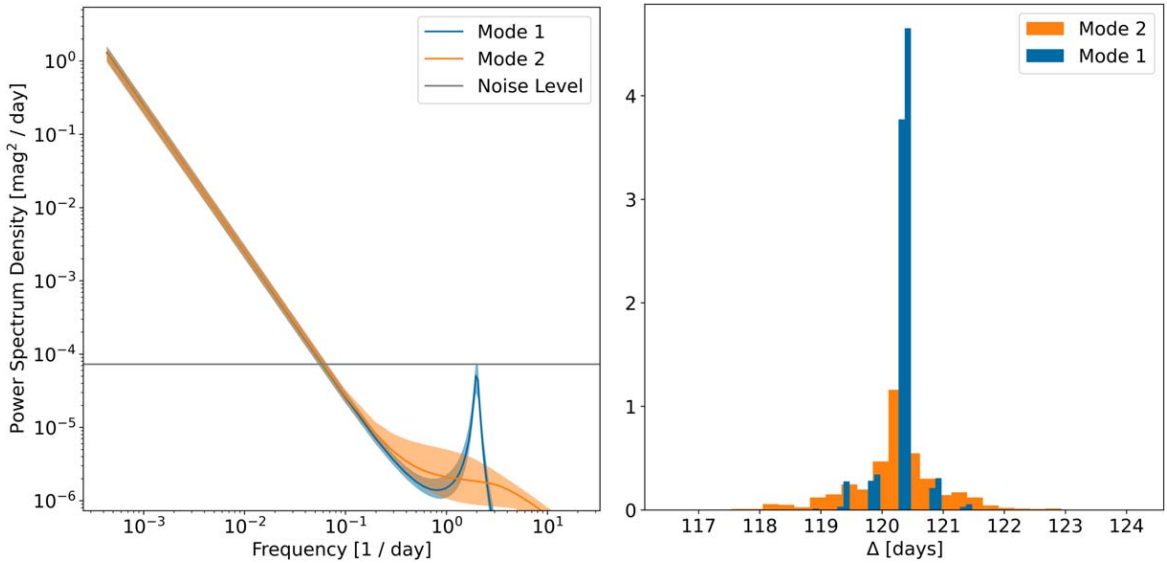

**Figure 2.** Left: a Monte Carlo sample from the posterior distribution for the PSD of the intrinsic light curve for quasar SDSS J1001+5027 under TD-CARMA (3, 1, 3). The posterior distribution for the CARMA parameters $\alpha$, $\beta$, $\sigma$ is bimodal, with one mode corresponding to a PSD with a frequency around 2 days (mode 1), and the other without a frequency. The frequency in mode 1 is below the measurement noise level. Right: histogram of posterior draws for $\Delta$, for each of the two modes in the joint posterior distribution computed with TD-CARMA (3, 1, 3), for the SDSS J1001+5027 data. For mode 1, since a frequency of 2 (day$^{-1}$) is detected, this constrains $\Delta$ to values that are multiple of 0.5 days, and thus drastically reduces the uncertainty. The standard deviation of the posterior of $\Delta$ in mode 2 is 0.686 days, and is 0.224 days in mode 1.

## 4. Simulated Data Analyses

In this section, we generate synthetic data and study the prediction accuracy and coverage probability of the estimates and credible intervals obtained for $\Delta$ with TD-CARMA. The primary purpose of this simulation study is to assess the reliability of the credible intervals and error bars on $\Delta$ derived by TD-CARMA ahead of the quasar data analyses presented in Section 5.

### 4.1. Simulation Design

We simulate doubly lensed synthetic data sets under nine different data generation models. The nine models were specified by three different continuous time series processes: DRW (i.e., CARMA(1, 0)), CARMA(2, 1), and CARMA(3, 1) crossed with three different orders of polynomial microlensing: $m = 1, 2, 3$. For each of the nine resulting data generating processes for the microlensed light curves, 300 replicate data sets are simulated, sharing all model parameters $\Omega$ related to the latent process. This yields 2700 data sets in total. In the next few paragraphs, we detail our procedure for the simulation of the synthetic data sets which, in an effort to mimic potential light curves from the Rubin Observatory Legacy Survey of Space and Time,[11] roughly follows the simulation design presented in Liao et al. (2015).

All 2700 data sets share the same generative values for $\Delta$, $\theta$, and $\mu$. We generate the shared value of $\Delta$ by sampling from a uniform distribution on [0, 200] days, and sample $\theta$ and $\mu$ from their priors, defined in Section 3.2. We then generate the time series parameters $\Omega$ for the three different intrinsic light-curve models DRW, CARMA(2, 1), and CARMA(3, 1). When an intrinsic light curve is generated under a DRW process, we set the timescale $\tau$ to 1000 days. For the CARMA(2, 1), we choose $\alpha$ such that the auto-correlation function (ACF) of the process features an exponentially damped sinusoid with a frequency equal to 1/100 days and an $e$-folding timescale of 1/500 days. For the CARMA(3, 1) process, we modify the $\alpha$ generated for the CARMA(2, 1) process to add an exponential decay with an $e$-folding timescale of 1/1000 days to the ACF of the CARMA (2, 1) process.[12] The moving average parameter vector $\beta$ is sampled uniformly from its prior, defined in Section 3.2. Finally, the $\sigma^2$ parameter is computed such that all three intrinsic light curves share the same marginal variance, $R(0)$; expressions for $R(0)$ are given in Equations (A1) and (A5) in Section A.2. We sample this shared value of $R(0)$ uniformly from the [0, 10] range. Figure 7 shows the ACFs (defined in Sections A.1 and A.2) for the three intrinsic light-curve models resulting from this procedure.

Given the generative values of the parameters $\Omega$, $\Delta$, and $\theta$, we create a doubly lensed synthetic data set as follows.

1. Given $\Omega$, generate an observation of the intrinsic light curve $Z(t)$ in a high-resolution grid of measurement times, with 100,000 evenly spaced epochs over a 1000 day period.
2. Given the values of $\Delta$ and $\theta$, generate high-resolution observations of the light curves $X(t)$ and $Y(t)$, via Equation (3).
3. Downsample the data to five observational seasons of length four months, and to a median observation cadence of three days. This yields approximately 200 data points for each light curve. The seasonal thinning is the same across all data set replicates, but the daily thinning is random.
4. Generate measurement error standard deviations $\{\delta_i^x\}_{i=1}^n$ and $\{\delta_i^y\}_{i=1}^n$, sampled from a normal distribution with mean and standard deviation equal to $10\%\sqrt{R(0)}$. The measurements $\{x_i\}_{i=1}^n$ and $\{y_i\}_{i=1}^n$ are then polluted with noise, and sampled at each observation time from the normal distributions $N(0, \delta_i^{x2})$ and $N(0, \delta_i^{y2})$.

---
[11] https://www.lsst.org/about

[12] The reader is referred to Kelly et al. (2014) for a description of the transformation from PSD frequencies and timescales to CARMA parameters $\alpha$.

For every generated data set, we fit nine different time delay models: TD-DRW, TD-CARMA(2, 1), and TD-CARMA(3, 1), each crossed with $m = \{1, 2, 3\}$.

### 4.2. Fidelity Metrics

We use two different classes of metrics to evaluate the performance of TD-CARMA in the context of this simulation study. First, we assess TD-CARMA purely in terms of its statistical estimator of $\Delta$ and use the classical measures of statistical fidelity of bias, variance, and root-mean-squared error (RMSE). Roughly speaking, these metrics describe the statistical properties of the error $\hat{\Delta} - \Delta$ of the estimator $\hat{\Delta}$ (see Appendix B.1 for a rigorous definition).

To facilitate comparison with other time delay methods developed in the astrophysics literature, we also evaluate TD-CARMA as a time delay estimation technique using the accuracy, precision, and $\chi^2$ metrics defined in Liao et al. (2015) for their TDC.

Specifically, suppose $N$ simulated doubly lensed data sets with true time delays $\{\Delta_i\}_{i=1}^N$ are fitted with TD-CARMA, which produces estimates $\{\hat{\Delta}_i\}_{i=1}^N$ and posterior standard deviations $\{\sigma_i\}_{i=1}^N$. Then the accuracy $A$, precision $P$, and $\chi^2$ are defined as

$$A = \frac{1}{N}\sum_{i=1}^{N} \frac{\hat{\Delta}_i - \Delta_i}{\Delta_i}, \quad (14)$$

$$P = \frac{1}{N}\sum_{i=1}^{N} \frac{\sigma_i}{\Delta_i}, \quad (15)$$

$$\chi^2 = \frac{1}{N}\sum_{i=1}^{N} \left(\frac{\hat{\Delta}_i - \Delta_i}{\sigma_i}\right)^2, \quad (16)$$

respectively.

### 4.3. Results

We fit the nine different inference models (TD-DRW($m$) and TD-CARMA($p$, $q$, $m$) with $m = \{1, 2, 3\}$, $p = \{2, 3\}$, and $q = 1$) to each of the 2700 generated data sets, and for every data set we select the inference model with the highest Bayesian evidence for our estimate of the time delay. We then evaluate the performance of the 300 selected estimates for each of the nine data generation models using the metrics described in Section 4.2 and report them in Table 2. The results indicate that the highest evidence models (the overwhelming majority of which are TD-CARMA(3, 1, ·)) are able to estimate the true value of $\Delta$ to very high accuracy and precision. Irrespective of the generative model, the (absolute) accuracy of TD-CARMA is $|A| = 0.00016$ and precision is $P = 0.00358$, and the chi-squared metric is $\chi^2 = 0.63669$. For comparison, entries in the 2015 TDC (Liao et al. 2015) were considered competitive if $|A| < 0.03$, $P < 0.01$, and $\chi^2 < 1.5$. The bias, variance, and RMSE metric also illustrate the quality and robustness of the TD-CARMA estimator for time delay $\Delta$. Out of the 2700 data sets, the TD-DRW model was only selected five times as the highest evidence model, which clearly shows the lack of flexibility of TD-DRW in modeling complex lens systems.

### 4.4. Robustness and Consistency Checks

To ensure that our error bars are well calibrated, we examine the frequency coverage of the posterior intervals. Specifically, we examine the proportion of simulated data sets for which the posterior intervals contain the true value of $\Delta$, i.e., the coverage probability of the intervals. To be well calibrated from a frequency point of view, posterior intervals should also be confidence intervals (as defined in Equation (B4)), e.g., a (say) 95% confidence interval for a model parameter is defined to be an interval computed from the data in such a way that upon repeated sampling of the data set (and thus repeated computation of the interval), at least 95% of the intervals contain the true value of the parameter. We therefore aim to check that posterior intervals with posterior probability $X\%$ (i.e., the theoretical coverage) have (approximately) a coverage probability of $X\%$ (i.e., the actual coverage). The reader is referred to Appendix B for a review of frequency coverage analyses.

Figure 3 plots the theoretical coverage against the actual coverage probability observed in our simulation study, for $p_n \in [0, 1]$. The actual coverage is the proportion of $p_n \times 100\%$ intervals (obtained from the highest evidence inference model for each of the 2700 simulated data sets) that actually contained the true value of $\Delta$.

The coverage of the 66% ($1\sigma$) and 95% ($2\sigma$) intervals, reported in Table 2 for each of the nine generative models, indicates overcoverage at the 66% level and good coverage at the 95% level. This is corroborated by Figure 3, where the estimated coverage line lies slightly above the theoretical coverage (45° line). This means that TD-CARMA produces conservative posterior intervals (i.e., slightly overestimated error bars), a desirable property in the context of time delay estimation. Nevertheless, our analyses of COSMOGRAIL quasar data sets in Section 5 show that TD-CARMA still produces much tighter error bars than other methods in the literature.

In conclusion, this simulation study demonstrates the flexibility of TD-CARMA in modeling lens systems of varying complexity, and the very high accuracy and precision at which TD-CARMA recovers the time delay parameter, $\Delta$.

## 5. Analysis of Six Lensed Quasars from COSMOGRAIL

### 5.1. Overview of the Numerical Analysis

In this section, we apply TD-CARMA to a selection of doubly lensed quasar data sets, and compare the resulting estimates to those obtained in the literature. We have selected six quasars that appear to have particularly complex intrinsic light curves and on which TD-DRW failed to return reliable estimates of $\Delta$ (using the *timedelay* package of Tak et al. 2017). Thus, a DRW process may not be sufficiently robust to model the underlying light curves.

For each data set we fit a total of 28 models. Specifically, we fit TD-DRW with $m = 3$ and TD-CARMA with auto-regressive order $p = \{2, 3, 4\}$, moving average order $q = \{1, 2, 3\}$, and microlensing order $m = \{1, 2, 3\}$, such that $p > q$. In Appendix C, we report the full results of our analyses in tables that give summary statistics for the posterior distribution of the model parameters for each fitted model, including the posterior mean and variance for $\Delta$, the Bayesian (log) evidence for the model, and the posterior mean of the frequency (computed from the posterior mean of the auto-regressive parameters $\alpha$) if the inferred PSD features a frequency (or frequencies). In the

Table 2
Fidelity Metrics Obtained by Selecting the Inference Model with the Highest Bayesian Evidence for Each Simulated Data Set, Grouped by Data Generation Model (DRW($m$) and CARMA($p, q, m$)*

| Metric | Data Generation Model | | | | | | | | | |
|---|---|---|---|---|---|---|---|---|---|---|
| | DRW(1) | DRW(2) | DRW(3) | C(2,1,1) | C(2,1,2) | C(2,1,3) | C(3,1,1) | C(3,1,2) | C(3,1,3) | Average |
| Bias | −0.00643 | −0.00437 | −0.00812 | −0.01099 | −0.01144 | −0.01618 | −0.00223 | 0.00026 | 0.01877 | 0.00452 |
| Variance | 0.00589 | 0.00622 | 0.00496 | 0.00164 | 0.00187 | 0.00250 | 0.00212 | 0.00208 | 0.03592 | 0.00702 |
| RMSE | 0.00593 | 0.00624 | 0.00503 | 0.00176 | 0.00201 | 0.00277 | 0.00213 | 0.00208 | 0.03628 | 0.00713 |
| Accuracy | −0.00022 | −0.00015 | −0.00028 | −0.00038 | −0.00040 | −0.00056 | −0.00008 | 0.00001 | 0.00065 | −0.00016 |
| Precision | 0.00321 | 0.00335 | 0.00312 | 0.00182 | 0.00178 | 0.00168 | 0.00543 | 0.00596 | 0.00594 | 0.00358 |
| $\chi^2$ | 0.63669 | 0.64569 | 0.63379 | 0.63285 | 0.72019 | 0.94437 | 0.09517 | 0.06484 | 1.24116 | 0.62386 |
| 1$\sigma$ Coverage[a] | 0.77 | 0.78 | 0.76 | 0.76 | 0.72 | 0.56 | 0.99 | 0.99 | 0.71 | 0.78 |
| 2$\sigma$ Coverage[b] | 0.99 | 0.98 | 0.93 | 0.99 | 0.97 | 0.84 | 1.0 | 1.0 | 0.89 | 0.96 |

**Notes.** * Abbreviated as C($p, q, m$) in the table.
[a] The theoretical coverage of a 1$\sigma$ interval is 66%.
[b] The theoretical coverage of a 2$\sigma$ interval is 95%.

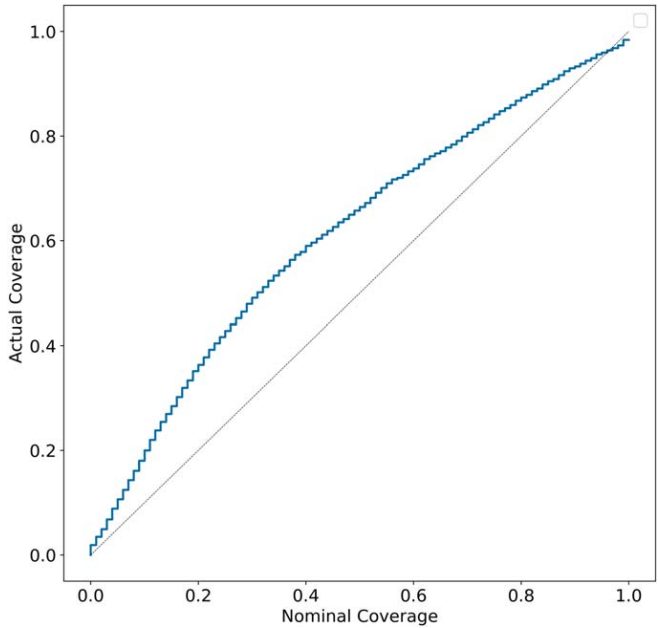

**Figure 3.** Nominal and actual coverage of generative value of $\Delta$ by posterior intervals produced by TD-DRW, TD-CARMA (2, 1), and TD-CARMA (3, 1) in the simulation study of Section 4, averaged over the three generative models. The dotted gray line is the $y = x$ line where the actual coverage equals the theoretical coverage.

event of a multimodal posterior distribution, we report these statistics for each mode and give the value for their Bayesian local (log) evidence, i.e., the mode-specific evidence.[13]

In all cases where the inferred PSD featured a frequency, the associated posterior distribution for the power of the frequency was below the measurement noise level. Therefore, following Section 3.5, we ignore all fits that included a frequency. We elaborate the numerical results for the frequencies analyses for all six examples in Section 5.6.

We select the model with highest Bayesian evidence (or relative probability), and compare its estimates for $\Delta$ with those derived by a baseline model, specifically our simplest model, TD-CARMA(2, 1, 3),[14] and those obtained by other methods in the literature. If the posterior distribution of the CARMA parameters under a particular model contains a mode with a frequency, we discard this mode and report the Bayesian evidence obtained by summing over the remaining modes (without frequencies). If the joint posterior is multimodal in the CARMA parameters, but none of the modes correspond to a frequency, we simply report the full Bayesian evidence given in Equation (11). The final estimates of $\Delta$ are reported in Table 3.

---

[13] In all of the applications, the TD-DRW model produced highly multimodal posterior distributions for $\Delta$ (10+ modes). For convenience, we therefore omit TD-DRW from the tables given in the Appendix, but report the relevant results in the main text.

### 5.2. Application I: HS2209+1914

HS2209+1914 is a doubly lensed quasar observed as part of the COSMOGRAIL study (Eulaers et al. 2013) with four telescopes: Euler, Mercator, Maidanak, and HCT, during a total of 8.5 years (2004 to mid-2012). Eulaers et al. (2013) propose a time delay measurement of $\Delta = -20 \pm 5$ days (using a combination of four different techniques, referenced in Table 4), warning that the lack of obvious fast variability features in the observed light curves and the possible presence of microlensing made it a particularly difficult observation to study.

Our analysis also suggests the presence of a strong microlensing effect in this observation, as indicated by the dependence of the estimated time delay, $\hat{\Delta}$, on the microlensing order, $m$. TD-CARMA models with $m = \{1, 2\}$ microlensing produce estimates of $\Delta$ around $-40$ days, whereas those with $m = 3$ produce $\hat{\Delta} \approx -22$ days (see Table 11); the latter is consistent with the literature (Eulaers et al. 2013; Kumar et al. 2015). (Our fitted $\hat{\Delta}$ for all models that did not detect a frequency appear in Table 11 in Appendix C.) To conduct a more principled model comparison when models indicate different possibilities of $\Delta$, we use the values of $\ln(\mathcal{Z})$ computed by MultiNest. This allows us to quantify the evidence in favor of models with $m = 3$, and the posterior probability of each model, via Equation (12). The cumulative probability of the $m = 3$ models is 0.999.

Our implementation of TD-DRW using MultiNest produced a (joint) posterior distribution with a total of 10 modes, and modes for $\Delta$ spanning from $\approx -11$ days to $\approx 43$ days. The bulk of the probability mass of the posterior of $\Delta$ under TD-DRW is

---

[14] We choose $m = 3$ for our baseline model to ensure that it is sufficiently flexible to capture the potential effect of microlensing. Moreno et al. (2019) notably supported the choice of CARMA(2, 1) to model AGN light curves.

**Table 3**
Summary of our Analysis (Carried out in Section 5, of Six Doubly Lensed Quasars from SDSS, Observed and Previously Analyzed by the COSMOGRAIL Collaboration)

| Data Set | COSMOGRAIL reference | COSMOGRAIL estimate | TD-CARMA |
|---|---|---|---|
| SDSS HS2209+1914 | Eulaers et al. (2013) | $-20 \pm 5$ | $-21.96 \pm 1.448$ |
| SDSS J1001+5028 | Kumar et al. (2013) | $119.1 \pm 3.3$ | $120.93 \pm 1.015$ |
| SDSS J1206+4332 | Millon et al. (2020b) | $113.0 \pm 3$ | $111.51 \pm 1.452$ |
| SDSS J1515+1511 | Millon et al. (2020b) | $210.5^{+5.5}_{-5.7}$ | $210.80 \pm 2.18$ |
| SDSS J1455+1447 | Millon et al. (2020b) | $47.2^{+7.5}_{-7.8}$ | $45.36 \pm 1.93$ |
| SDSS J1349+1227 | Millon et al. (2020b) | * | $432.05 \pm 1.950$ |

**Table 4**
Comparison of Our Highest Evidence Model, i.e., TD-CARMA(3, 2, 3), and our Baseline, TD-CARMA(2, 1, 3), to Estimates of $\Delta$ for HS2209+1914 in the Literature

| Technique | $\hat{\Delta}$ | SD($\hat{\Delta}$) |
|---|---|---|
| Dispersion[a] | $-20.99$ | 10.09 |
| Difference[a] | $-20.08$ | 13.78 |
| Spline[a] | $-19.77$ | 6.03 |
| NMF[a] | $-19.28$ | 1.48 |
| *Combination of the four above*[a] | $-20$ | 5 |
| Difference smoothing (modified)[b] | $-22.9$ | 5.3 |
| ROA[c] | $-24.0$ | 2.4 |
| TD-CARMA(2, 1, 3) | $-21.74$ | 1.423 |
| TD-CARMA(3, 2, 3) | $-21.96$ | 1.448 |

**Notes.**
[a] Eulaers et al. (2013), where estimates for $\Delta$ are reported without the minus sign, as they swap labels for the two light curves compared to our notation.
[b] Kumar et al. (2015).
[c] Donnan et al. (2021).

**Table 5**
Posterior Mean and Standard Deviation for $\Delta$ under the Models with the Highest Bayesian Log-evidence, with Relative Probability $\pi$ Summing to 0.999, for HS2209+1914

| Model | $\hat{\Delta}$ | SD($\hat{\Delta}$) | $\ln(\mathcal{Z})$ | $\pi$ |
|---|---|---|---|---|
| TD-DRW(3) | 20.23 | 0.918 | 2536.00 | 0.000 |
| TD-CARMA (3, 2, 3) | $-21.96$ | 1.448 | 2760.24 | 0.601 |
| TD-CARMA (4, 2, 3) | $-21.95$ | 1.403 | 2759.83 | 0.399 |

around ≈20 days, which indicates that TD-DRW fails even to get the sign of $\Delta$ correct. In addition, Tables 5, 11, and 12 show the Bayesian evidence $\ln(\mathcal{Z})$ in favor of TD-DRW is the lowest of all the considered models.

The top left panel of Figure 4 compares the posterior distributions for $\Delta$ under the TD-CARMA model with the highest evidence, i.e., TD-CARMA(3, 2, 3), and the baseline model, i.e., TD-CARMA(2, 1, 3), with estimates found in the astrophysics literature. Figure 5 plots the data against the model predictions derived by TD-CARMA(3, 2, 3). Eulaers et al. (2013) propose an estimate of $\hat{\Delta} = -20 \pm 5$ days, and Kumar et al. (2015) give $\hat{\Delta} = -22.9 \pm 5.3$ days. TD-CARMA(3, 2, 3) and TD-CARMA(2, 1, 3) estimate $\hat{\Delta} = -21.96 \pm 1.448$ and $\hat{\Delta} = -21.74 \pm 1.423$, hence significantly reducing the error bars relative to existing methods (see Table 4). The simulation study in Section 4 demonstrates that our error bars are reliable and conservative.

### 5.3. Application II: SDSS J1001+5027

SDSS J1001+5027 is a doubly lensed quasar, observed by the COSMOGRAIL collaboration (Kumar et al. 2013) with three telescopes: Mercator, Maidanak, and HCT for more than six years (from 2005 March to 2011 July), producing 443 observational epochs.

Our implementation of TD-DRW using MultiNest produced a (joint) posterior distribution with a total of 20 modes, and modes for $\Delta$ spanning values from ≈123 days and ≈135 days. The bulk of the probability mass of the posterior of $\Delta$ is around ≈132 days, almost two weeks away from the consensus value in the relevant literature, which is ≈120 days. The TD-DRW(3) model has $\ln(\mathcal{Z}) = 1803.30$, well below that of any of the TD-CARMA($p, q, m$) (see Tables 6 and 13).

Table 6 shows the five TD-CARMA($p, q, m$) with the highest Bayesian (log) evidences, as well as their relative probabilities, $\pi_i$, computed as detailed in Section 3.3. Together the five top models account for 97.7% of the total model probabilities. For given values of hyperparameters $p$ and $q$, the increase in $\ln(\mathcal{Z})$ obtained from increasing $m$ is small, indicating a weak effect of microlensing in this observation; see Tables 6 and 13 (the latter in Appendix C).

In Table 7, we compare the TD-CARMA($p, q, m$) model with the highest Bayesian (log) evidence, i.e., TD-CARMA(4, 3, 2), with recently published analyses of SDSS J1001+5027. TD-CARMA(4, 3, 2) produces error bars that are between 30% and 83.63% lower than the existing methods; the baseline model, i.e., TD-CARMA(2, 1, 3), produces error bars that are between 48.34% and 87.92% lower. The larger error bars of TD-CARMA(4, 3, 2) relative to TD-CARMA(2, 1, 3) may be attributable to a bias–variance trade off: more highly parameterized models are more flexible and tend to provide better fits, but at the cost of higher levels of uncertainty.

Kumar et al. (2013), the main reference in the estimation of the time delay of SDSS J1001+5027, argue that their four independent measurement techniques (listed in Table 7) each have their own limitations and thus combine the individual estimations from each method to produce their final estimate: $\hat{\Delta} = 119.3 \pm 3.3$ days. Our final estimate is $\hat{\Delta} = 120.93 \pm 1.015$ days, thus providing a consistent (since our posterior distribution covers their estimate) yet novel estimate of the time delay, that is three times more certain.

### 5.4. Application III: SDSS J1206+4332

SDSS J1206+4332 is a doubly lensed quasar observed by the COSMOGRAIL collaboration (Eulaers et al. 2013) with three telescopes: Maidanak, Mercator, and HCT for a total of seven years (from 2005 to 2011).

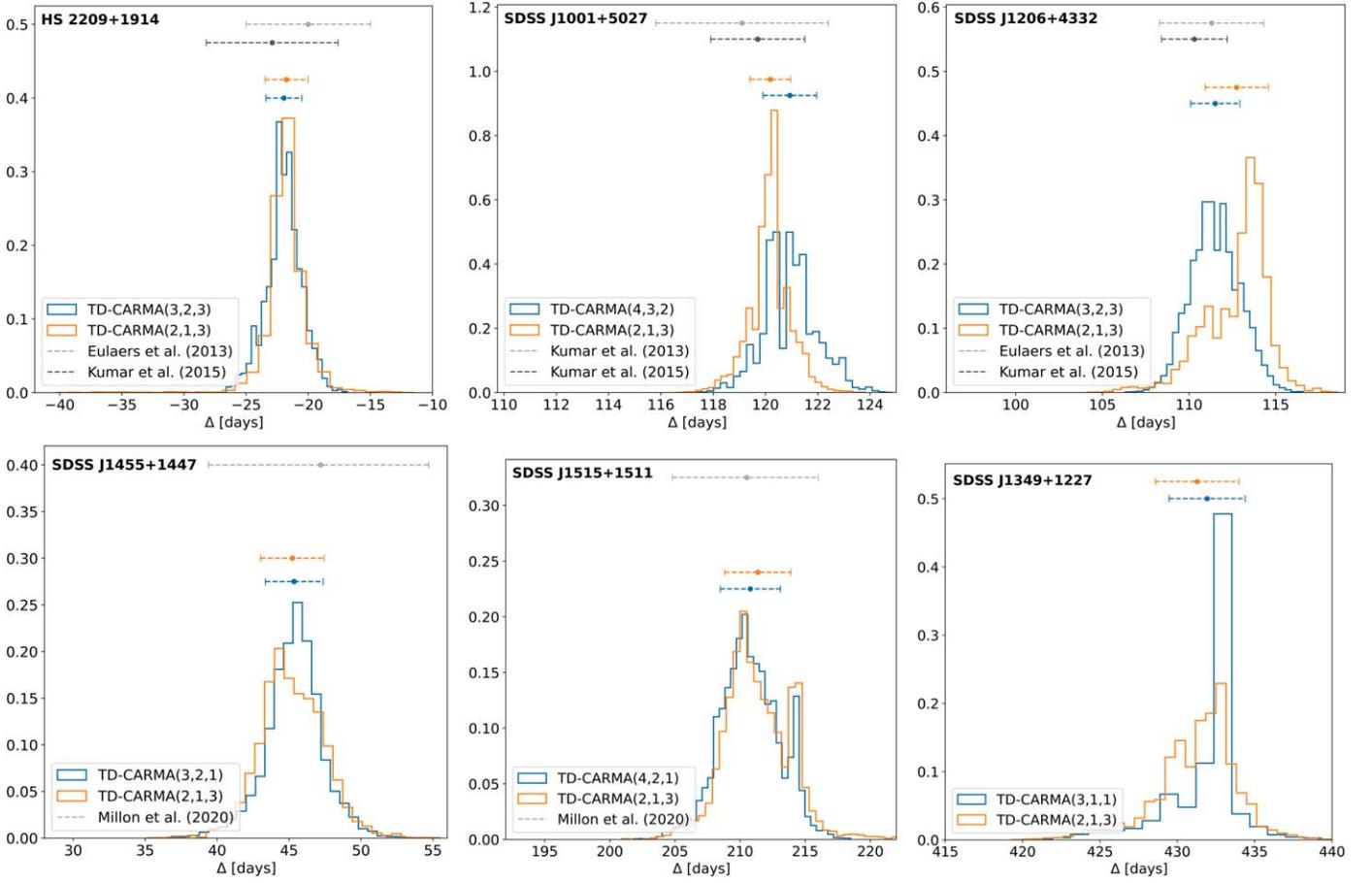

**Figure 4.** Point estimates and $1\sigma$ error bars obtained both by our highest evidence model and the baseline model, i.e., TD-CARMA(2, 1, 3), compared to the previously published methods in Eulaers et al. (2013), Kumar et al. (2013), and Millon et al. (2020b) from the COSMOGRAIL collaboration, and Kumar et al. (2015). In all six cases, both our simplest and best model obtain significantly tighter error bounds than those published in the abovementioned papers. On the HS2209+1914 doubly lensed quasar data set (top left panel), our methods obtain error bars between 71.04% and 73.15% smaller than Eulaers et al. (2013) and Kumar et al. (2015), corresponding to a reduction in the uncertainty by a factor of $\approx 3.5$. For the SDSS J1001+5027 data set (top middle panel), we achieve error bars between $\approx 1.8$ and $\approx 4.4$ times smaller than the existing analyses, (depending on which model we use and to which method we compare it). For SDSS J1206+4332 (top right panel), we reduce the error estimates by a factor of $\approx 1.12-2.06$; for SDSS J1455+1447 (bottom left panel), we reduce the error estimates by a factor of $\approx 4$; for SDSS J1515 +1511 (bottom middle panel), we reduce the error estimates by a factor of $\approx 2.5$. Millon et al. (2020b) could not produce an estimate on $\Delta$ for SDSS J1349+1227 (bottom right panel).

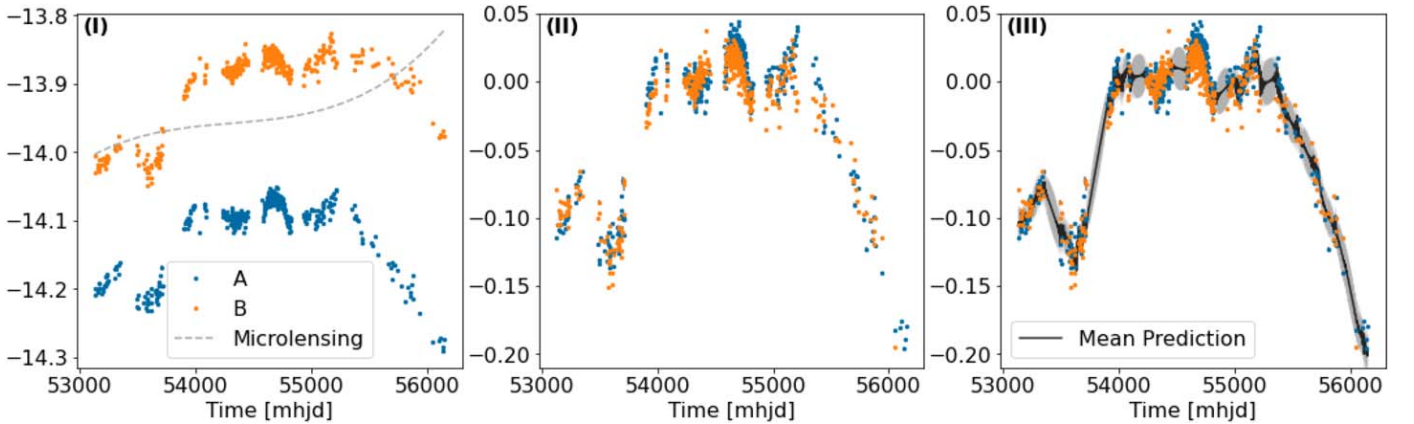

**Figure 5.** Panel (I) shows the data for the HS2209+1914 doubly lensed quasar, and the microlensing polynomial regression curve (with $m = 3$) fit to light curve B (corresponding to measurements $\boldsymbol{y}$ in our mathematical notation reported in Table 1). The effect of microlensing can be seen, for example in the time period between 55,500 and 56,000 MHJD, where light curve B features a slower decay in magnitude than light curve A. This is successfully captured and adjusted for by the increasing behavior in the fitted microlensing curve. Panel (II) shows the data in Panel (I), adjusted for time delay $\Delta$ and the microlensing effect modeled by the polynomial regression, which corresponds to the composite light curve $\boldsymbol{z}$ defined in Equation (4). Panel (III) shows the mean fitted values (and uncertainties) of the CARMA(3, 2) process to composite light curve $\boldsymbol{z}$.

**Table 6**
Posterior Mean and Standard Deviation for $\Delta$ under the Five Models with the Highest Bayesian Log-evidence, for SDSS J1001+5027

| Model | $\hat{\Delta}$ | SD($\hat{\Delta}$) | ln($\mathcal{Z}$) | $\pi$ |
|---|---|---|---|---|
| TD-DRW(3) | 132.71 | 0.770 | 1803.30 | 0.000 |
| TD-CARMA(4, 3, 2) | 120.93 | 1.015 | 2761.25 | 0.416 |
| TD-CARMA(4, 3, 3) | 120.21 | 1.018 | 2761.12 | 0.366 |
| TD-CARMA(3, 2, 3) | 120.16 | 1.003 | 2759.48 | 0.104 |
| TD-CARMA(3, 2, 2) | 120.88 | 1.013 | 2759.44 | 0.068 |
| TD-CARMA(4, 2, 3) | 120.21 | 0.970 | 2758.35 | 0.023 |

**Table 7**
Comparison of Our Highest Evidence Model, i.e., TD-CARMA(4, 3, 2), and our Baseline, TD-CARMA (2, 1, 3), to Estimates of $\Delta$ for SDSS J1001+5027 in the Literature

| Technique | $\hat{\Delta}$ | SD($\hat{\Delta}$) |
|---|---|---|
| Dispersion-like[a] | 120.5 | 6.2 |
| Difference smoothing[a] | 118.6 | 3.7 |
| Regression difference[a] | 121.1 | 3.8 |
| Free-knot spline[a] | 119.7 | 2.6 |
| *Combination of the four above*[a] | 119.1 | 3.3 |
| GPs[b] | 117.8 | 3.2 |
| Difference smoothing (modified)[c] | 119.7 | 1.8 |
| ROA[d] | 119.9 | (−1.5, 1.4) |
| TD-CARMA(2, 1, 3) | 120.18 | 0.749 |
| TD-CARMA(4, 3, 2) | 120.93 | 1.015 |

**Note.**
[a] Eulaers et al. (2013).
[b] Hojjati et al. (2013).
[c] Kumar et al. (2015).
[d] Donnan et al. (2021).

For this data, TD-DRW has the lowest Bayesian evidence of the set of considered models, and has a relative probability of essentially 0 (up to numerical precision). TD-DRW produces a joint posterior distribution with six modes, with posterior modes for $\Delta$ ranging from $\approx$106 days to $\approx$117 days. Even though our computational procedure allows us to obtain a sample from the posterior (whereas the MCMC based implementation of TD-DRW in the *timedelay* package from Tak et al. 2017 fails to converge), TD-DRW does not appear to be sufficiently flexible for this data set.

Turning to fits obtained with TD-CARMA, Table 8 shows that the posterior distribution of $\Delta$ is sensitive to the value of $m$, which indicates the presence of a strong microlensing effect in this observation. Indeed, Figure 6 illustrates that third-order polynomial regression seems to pull the posterior distribution of $\Delta$ toward $\approx$112 days, whereas the posterior distributions of $\Delta$ obtained with $m = 2$ tend to center around $\approx$109 days. The posterior PSDs of the intrinsic light curve under both models are identical, which shows that this effect is indeed solely attributable to microlensing. In some cases, such as TD-CARMA(2, 1, 3), the posterior distribution covers both areas of likely values. This is also the case in Eulaers et al. (2013), where the numerical model fit (NMF) technique estimates $\hat{\Delta} = 113.80 \pm 0.80$, whereas the spline technique finds $\hat{\Delta} = 111.31 \pm 3.93$. This suggests that either the observational sampling does not allow a strong constraint for $\Delta$, or that the effect of microlensing may be too complex for the currently employed modeling approaches.

**Table 8**
Posterior Mean and Standard Deviation for $\Delta$ under the Five Models with the Highest Bayesian Log-evidence, for SDSS J1206+4332

| Model | $\hat{\Delta}$ | SD($\hat{\Delta}$) | ln($\mathcal{Z}$) | $\pi$ |
|---|---|---|---|---|
| TD-DRW(3) | 114.68 | 1.413 | 1551.02 | 0.000 |
| TD-CARMA(3, 2, 3) | 111.51 | 1.452 | 1639.06 | 0.999 |
| TD-CARMA(3, 2, 2) | 109.27 | 1.351 | 1628.43 | 2.42e-5 |
| TD-CARMA(2, 1, 3) | 112.78 | 1.693 | 1624.65 | 5.52e-7 |
| TD-CARMA(3, 1, 3) | 112.77 | 1.670 | 1623.66 | 2.05e-7 |
| TD-CARMA(2, 1, 2) | 108.33 | 2.240 | 1616.64 | 1.83e-10 |

TD-CARMA(3, 2, 3) dominates the model space, with probability 0.999, and estimates $\hat{\Delta} = 111.51 \pm 1.452$ days. This halves the uncertainty of Eulaers et al. (2013; $111.3 \pm 3$ days), and reduces the uncertainty derived in Kumar et al. (2015) by 23.58% ($110.3 \pm 1.9$ days), as shown in Table 9. Although the estimate produced by Donnan et al. (2021; see "ROA Method" in Table 9) has less uncertainty than TD-CARMA for this particular data set, its similarity to the results obtained by the NMF technique (Eulaers et al. 2013) suggest that it may have left part of the parameter space unexplored.

### 5.5. Three Applications from Millon et al. (2020b)

In this section, we apply TD-CARMA to three doubly lensed quasars observed as part of the COSMOGRAIL study and recently analyzed in Millon et al. (2020b): SDSS J1515+1511, SDSS J1455+1447, and SDSS J1349+1227. For SDSS J1515 +1511 and SDSS J1455+1447, we obtain estimates that are highly consistent with Millon et al. (2020b) and the relevant literature, and dramatically improve on the previously published uncertainties.

For SDSS J1515+1511, Shalyapin & Goicoechea (2017) propose a time delay measurement of $\Delta = 211 \pm 5$ days, and more recently Millon et al. (2020b) estimate $\Delta = 210.5^{+5.5}_{-5.7}$ days. Our model with highest evidence, TD-CARMA(4, 2, 1), estimates $\Delta = 210.80 \pm 2.18$ days, dramatically reducing the uncertainty compared to previously published estimates. The full results are reported in Tables 17 and 18 in Appendix C.

For SDSS J1455+1447, Millon et al. (2020b) propose a time delay estimate of $\Delta = 47.2^{+7.5}_{-7.8}$ days. Our highest evidence model, TD-CARMA(3, 2, 1), estimates $\Delta = 45.36 \pm 1.93$ days. Both our estimate and the one proposed on Millon et al. (2020b) are within one standard deviation of their respective error bars. Our estimate, however, is roughly four times less uncertain. The full results are reported in Tables 19 and 20 in Appendix C.

For SDSS J1349+1227, Kayo et al. (2010) predicted a time delay of the order of one year, and Millon et al. (2020b) could not determine the time delay of this lens system. We produced an estimate that is consistent with the one year order of magnitude value predicted in Kayo et al. (2010). For all values of the ($p$, $q$, $m$) hyperparameter triplet, TD-CARMA produces highly consistent estimates for $\Delta$, all gravitating around 431–432 days. Our highest evidence model, TD-CARMA(3, 1, 1), estimates $\Delta = 432.05 \pm 1.950$ days. The full results are reported in Tables 21 and 22 in Appendix C.

As with Applications I–III, the TD-DRW(3) model produced highly multimodal posterior distributions for $\Delta$. These estimates had much lower Bayesian evidence that our more

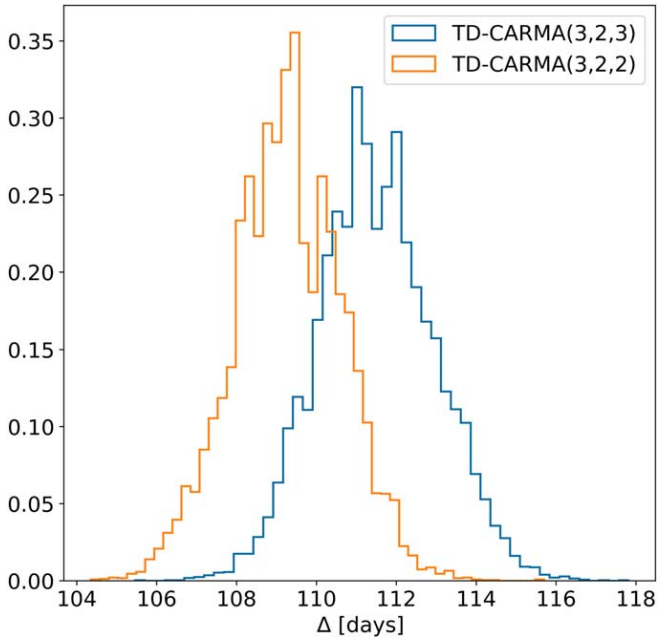

**Figure 6.** Histograms of posterior draws for $\Delta$ under TD-CARMA(3, 2, 2) and TD-CARMA(3, 2, 3), for the doubly lensed quasar data set of SDSS J1206+4332.

**Table 9**
Comparison of our Highest Evidence Model, i.e., TD-CARMA(3, 2, 3), and our Baseline, TD-CARMA(2, 1, 3), to Estimates of $\Delta$ for SDSS J1206+4332 in the Literature

| Technique | $\hat{\Delta}$ | SD($\hat{\Delta}$) |
| --- | --- | --- |
| Dispersion[a] | 113.65 | 6.79 |
| Difference[a] | 109.73 | 8.28 |
| Spline[a] | 111.31 | 3.93 |
| NMF[a] | 113.80 | 0.90 |
| *Combination of the four above*[a] | 111.3 | 3 |
| Splines+Regression difference[b] | 111.8 | (−2.7, 2.4) |
| Difference smoothing (modified)[c] | 110.3 | 1.9 |
| ROA[d] | 113.0 | (−1.2, 1.1) |
| TD-CARMA(2, 1, 3) | 112.78 | 1.693 |
| TD-CARMA(3, 2, 3) | 111.51 | 1.452 |

**Notes.**
[a] Eulaers et al. (2013).
[b] Birrer et al. (2019).
[c] Kumar et al. (2015).
[d] Donnan et al. (2021).

complex TD-CARMA models, and were inconsistent with previously published analyses.

### 5.6. Numerical Evaluation of Frequencies

Tables 12, 14, 16, 18, 20, and 22 in Appendix C show that under some of the TD-CARMA($p, q, m$), the posterior distribution of the CARMA parameters ($\alpha, \beta, \sigma$) correspond to PSDs with a frequency (i.e., a Lorentzian centered away from 0), either for specific modes or for the whole distribution. For instance, one mode of the posterior PSD under TD-CARMA(3, 1, 1) fitted to the SDSS J1206+4332 data set has a frequency of ≈1 day, and the posterior PSD under TD-CARMA(4, 1, 3) fitted to the SDSS J001+5027 data set finds a frequency of ≈2 days.

**Table 10**
Comparing the (Posterior) Power of the Detected Frequencies with the Measurement Noise Level

| Data Set | Frequency | 99% power[a] | Noise level[a] |
| --- | --- | --- | --- |
| J1206 | ≈1 days | [1.69, 2.94] | 10.67 |
| J1001 | ≈2 days | [0.22, 0.92] | 0.60 |
| HS2209 | ≈1 days | [1.28, 3.94] | 2.29 |

**Note.**
[a] Values are given ($\times 10^{-4}$).

These frequencies might be an inherent property of the intrinsic light curves themselves, or simply a feature of the measurement noise.

For illustration, Table 10 shows the (posterior mean of the) detected frequencies, the associated 99% posterior intervals for the power, and the measurement noise level for the SDSS HS2209+1914, SDSS J1001+5027, and SDSS J1206+4332 data sets. In all three cases, the 99% posterior intervals for the power of the detected frequencies either cover or are well below the measurement noise level.[15] We therefore discard all models that detect a frequency in our analyses.

## 6. Conclusion

We present TD-CARMA, an extension to TD-DRW presented by Tak et al. (2017) and designed to perform Bayesian estimation of cosmological time delays arising in gravitationally lensed systems of light curves. Because they are more flexible and more widely applicable for modeling the intrinsic stochastic variability of AGN light curves, we employ CARMA processes in our extension and thus design an accurate general method applicable to a wide range of AGN-lensed systems.

Our simulation studies and data applications demonstrate that our method is applicable and produces precise time delay estimates on data sets that cannot be analyzed by the DRW-based method proposed by Tak et al. (2017).

The computational aspects of our method also allow a number of improvements over existing methods. First, the method does not require a user-specified initial value of the time delay parameter $\Delta$. Our MultiNest-based Bayesian inference engine is robust to multimodal posterior distributions, which arise frequently both in CARMA processes and in time delay estimation problems. Finally, our method directly computes the Bayesian evidence for the fitted model. This allows us to perform principled Bayesian model comparison and selection, for instance to select hyperparameters including the order of the CARMA process and microlensing model.

There are several potential improvements and additional applications of TD-CARMA beyond what we present in this work. The models developed in this article only account for single-band light curves. Hu & Tak (2020) developed a method to model multiband light curves using the DRW process, and show that it yields more precise time delay estimates. An avenue for future research would be to extend our method to multiband light curves. Moreover, we carried out a prior sensitivity analysis and empirically verified the robustness of our model selection procedure to the choice of priors. Although we find that TD-CARMA estimates of $\Delta$ and their error bars are not particularly dependent on the choice of priors placed on the intrinsic light curve and microlensing parameters, the Bayesian evidence, $\mathcal{Z}$, tends to be more sensitive to the priors. Thus, the choice of model, based on $\mathcal{Z}$, can be more prior dependent. Insofar as the ultimate aim is to estimate the time delay, however, this is only a concern if there are competing models with appreciable posterior probabilities that deliver substantively different estimates of $\Delta$. If such a situation arises, these competing models can be combined via Bayesian model averaging (e.g., Hoeting et al. 1999). In our numerical examples, however, we either found a model that clearly dominated in terms of Bayesian evidence, or a small collection of models sharing most of the probability with very similar posterior distributions. In both cases, performing Bayesian model averaging would not lead to any major differences in our results.

---
[15] We chose to only include these three data sets in Table 10 for simplicity, as for each of them the detected frequencies among the different models were roughly the same. This is not the case for the data sets of SDSS J1515+1511, SDSS J1455+1447, and SDSS J1349+1227; however, we conducted the same analysis and reached the same conclusion as for the other three applications, namely that the 99% posterior intervals for the power of the detected frequencies cover or are below the measurement noise level.

Variants of TD-CARMA might also be applied to problems outside of time delay estimation. Time delays also arise in reverberation mapping, where the emission-line response is delayed with respect to changes in the continuum in accretion flows around supermassive black holes. The estimation of this delay can be used to estimate the mass of the black hole.

## Acknowledgments

This work was conducted under the auspices of the CHASC International Astrostatistics Center. CHASC is supported by NSF grants DMS-18-11308, DMS-18-11083, DMS-18-11661, DMS-21-13615, DMS-21-13397, and DMS-21-13605; and by the UK Engineering and Physical Sciences Research Council [EP/W015080/1]. We thank CHASC members for many helpful discussions, especially Vinay Kashyap and Kaisey Mandel, and we thank Andy Thomas for his continuous help and support on computational matters. A.D.M. is supported by the EPSRC Centre for Doctoral Training in Modern Statistics and Statistical Machine Learning (EP/S023151/1). D.v.D. and A.D.M. were also supported in part by a Marie-Skodowska-Curie RISE Grant (H2020-MSCA-RISE-2019-873089) provided by the European Commission. A.S. further acknowledges support from NASA contract to the Chandra X-ray Center NAS8-03060.

*Software:* TD-CARMA: https://github.com/antoinedmeyer/td_carma, MultiNest (Feroz et al. 2009; Buchner et al. 2014), EzTao (Yu & Richards 2022), and celerite (Foreman-Mackey et al. 2019).

The COSMOGRAIL data for the HS2209+1914, SDSS J1001+5027, SDSS J1206+4332, SDSS J1515+1511, SDSS J1455+1447, and SDSS J1349+1227 doubly lensed quasar data sets are publicly available online at https://obswww.unige.ch/~millon/d3cs/COSMOGRAIL_public/code.php.

## Appendix A
## Modeling Stochastic Variability in Astronomical Objects

To compare different classes of stochastic processes and to gain insight about the underlying process governing their variability, it is useful to inspect the ACF and PSD.

### A.1. The DRW Process

The ACF of a stochastic process $X(t)$, $t \in \mathbb{R}$ is defined as $E[(X(t_i) - E(X(t_i)))(X(t_j) - E(X(t_j)))]$, and gives the covariance of the process with itself at two different times. The ACF of a DRW process is given by

$$R(d_{ij}) = \frac{\sigma^2 \tau}{2} e^{-\frac{d_{ij}}{\tau}}, \quad (A1)$$

where $d_{ij} = t_i - t_j$ is the time difference between any two observation times $t_i$ and $t_j$, with $t_i > t_j$ (Kelly et al. 2009). The ACF in Equation (A1) features an exponentially decaying component with $e$-folding timescale $\tau$ which highlights the mean-reverting behavior of the DRW process.

The PSD, defined as the Fourier transform of the ACF, describes the distribution of power[16] over frequencies for continuous and stationary stochastic processes. The PSD of the DRW process, given by

$$P(f) = \sigma^2 \frac{1}{\left(\frac{1}{\tau}\right)^2 + (2\pi f)^2}, \quad (A2)$$

is a Lorentzian centered at 0, with a break frequency at $1/2\pi\tau^2$ (Kelly et al. 2009). This illustrates the limited applicability of a DRW process in describing the stochastic variability in light curves. For example, quasiperiodic variability features in an AGN light curve can be represented as Lorentzians centered away from 0; hence a DRW process would be unable to characterize this type of variability. Moreover, the DRW process can only account for a single break frequency in the PSD, which renders it somewhat inflexible.

### A.2. The CARMA Processes

To ensure stationarity of the process, we must have $q < p$, and the roots of the auto-regressive polynomial $A(z)$, defined as

$$A(z) = \sum_{k=0}^{p} \alpha_k z^k, \quad (A3)$$

must have negative real parts (Kelly et al. 2014; Tsai & Chan 2000; Tsai et al. 2011). In practice, this constraint is more easily enforced when considering an alternative parameterization of the auto-regressive polynomial (Jones 1981; Kelly et al. 2014), obtained by factorizing it into linear and quadratic terms as follows

$$A(z) = (a_1 + a_2 z + z^2)(a_3 + a_4 z + z^2)\ldots$$
$$\times \begin{cases} a_{p-1} + a_p z + z^2 & \text{if } p \text{ is even,} \\ a_p + z & \text{if } p \text{ is odd.} \end{cases}$$

The roots of the auto-regressive polynomial are negative if the coefficients $\boldsymbol{a} = \{a_1 \ldots, a_p\}$ are positive. We discuss computational strategies for stable implementation and inference in Section 3.4.

---

[16] The power of a stochastic process is defined as the squared value of the process, i.e., the power of process $X(t)$ is given by $|X(t)|^2$.

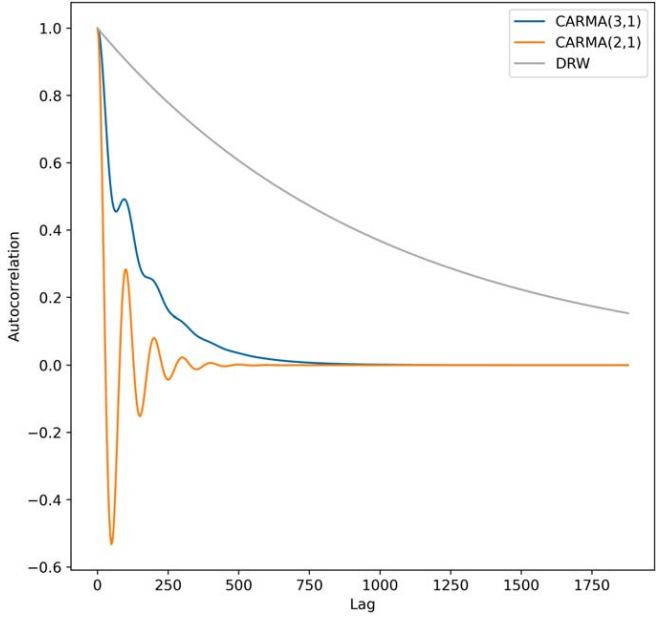

**Figure 7.** ACFs for the intrinsic light-curve models used in Simulation I.

The PSD of a stationary CARMA($p$, $q$) process is given by

$$P(f) = \sigma^2 \frac{\left|\sum_{j=0}^{q} \beta_j (2\pi i f)^j\right|^2}{\left|\sum_{k=0}^{p} \alpha_k (2\pi i f)^k\right|^2}, \quad (A4)$$

and its ACF is

$$R(d_{ij}) = \sigma^2 \sum_{k=1}^{p} \frac{\left[\sum_{l=0}^{q} \beta_l r_k^l\right]\left[\sum_{l=0}^{q} \beta_l (-r_k)^l\right] \exp(r_k d_{ij})}{-2\,\mathrm{Re}(r_k) \prod_{l=1, l\neq k} (r_l - r_k)(r_l^* + r_k)}, \quad (A5)$$

where $r_k$, $k = 1,..., p$ are the roots of the auto-regressive polynomial $A(z)$ (Tsai & Chan 2000; Tsai et al. 2011; Kelly et al. 2014).

The ACF of a CARMA process is therefore a weighted sum of exponentially damped sinusoidal functions (corresponding to complex roots) and exponential decays (corresponding to real roots; Kelly et al. 2014). A CARMA process has the flexibility to model exponentially decaying auto-correlations with multiple timescales (rather than a single one for the DRW process), and to incorporate periodic oscillations in the auto-correlations. Figure 7 illustrates the difference between the DRW and CARMA processes in terms of their auto-correlations and highlights the fact that even within a CARMA($p$, $q$) family, CARMA processes with different orders of $p$ and $q$ may look quite different. This implies that the flexibility of CARMA processes makes them suitable to characterize the variability in a wide range of AGN light curves.

## Appendix B
## Frequency Evaluation

This section provides some background on the frequentist analysis of Bayesian methods (following Tak et al. 2016), carried out in Section 4 of this paper. A key component of model evaluation is to verify that the confidence intervals for the parameter of interest have desired frequency properties.

In frequentist statistics, model parameters such as the time delay $\Delta$ are considered to be fixed, while the data are considered to be random; we denote random data by $\boldsymbol{D}$. A particular data set $\mathbf{d}$ is a random realization of $\boldsymbol{D}$. Our statistical model, TD-CARMA provides an estimator for $\Delta$, i.e TD-CARMA allows us to compute an estimate $\hat{\Delta}$ given observed data $\mathbf{d} \sim \boldsymbol{D}$.

### B.1. Frequency Evaluation of a Statistical Estimator

Here we formally define the concepts of statistical bias, variance, and RMSE used in Section 4 in our numerical evaluation of the statistical veracity of TD-CARMA. These quantities are widely used to describe the frequency properties of statistical estimators. Denote by $\hat{\Delta}|\boldsymbol{D}$ the distribution of the estimate for time delay parameter over replicates of data $\mathbf{d} \sim \boldsymbol{D}$, obtained by repeatedly fitting a given model to randomly generated replicate data sets. The bias of an estimate $\hat{\Delta}$ is defined as

$$\mathrm{Bias}(\hat{\Delta}) = \mathbb{E}_{\hat{\Delta}|\boldsymbol{D}}(\hat{\Delta}) - \Delta, \quad (B1)$$

i.e., the average estimate of $\Delta$ across replicate data sets, from which the true value of $\Delta$ is subtracted. Similarly, the variance of the estimate $\hat{\Delta}$ is defined as

$$\mathrm{Var}(\hat{\Delta}) = \mathbb{E}_{\hat{\Delta}|\boldsymbol{D}}[(\mathbb{E}_{\hat{\Delta}|\boldsymbol{D}}(\hat{\Delta}) - \hat{\Delta})^2]. \quad (B2)$$

Finally, the RMSE is defined as

$$\begin{aligned}\mathrm{RMSE}(\hat{\Delta}) &= \mathbb{E}_{\hat{\Delta}|\boldsymbol{D}}[(\hat{\Delta} - \Delta)^2] \\ &= \sqrt{\mathrm{Bias}(\hat{\Delta})^2 + \mathrm{Var}(\hat{\Delta})}.\end{aligned} \quad (B3)$$

Models with lower RMSEs are generally preferred as they return estimates that tend to be closer to the true value of the parameter.

### B.2. Frequency Coverage of Posterior Intervals

A $100(1-\alpha)\%$ confidence interval for parameter $\Delta$ is a random interval $[a(\boldsymbol{D}), b(\boldsymbol{D})]$ such that

$$\mathbb{P}(a(\boldsymbol{D}) < \Delta < b(\boldsymbol{D})) = 1 - \alpha. \quad (B4)$$

In other words, the interval $[a(\boldsymbol{D}), b(\boldsymbol{D})]$ covers $\Delta$ with probability $1 - \alpha$. In theory, if we were to sample new data $\mathbf{d} \sim \boldsymbol{D}$ repeatedly and construct a 95% confidence interval $[a(\mathbf{d}), b(\mathbf{d})]$ every time, then approximately 95% of our intervals will contain the true value of $\Delta$. The proportion of $100(1-\alpha)\%$ confidence intervals that contain the true value of $\Delta$ is referred to as the empirical or actual coverage of the $100(1-\alpha)\%$ confidence intervals, and $100(1-\alpha)\%$ is referred to as the theoretical coverage. A model with perfectly calibrated confidence intervals (i.e., error bars) would have its actual coverage equal to the theoretical coverage for all $\alpha \in [0, 1]$. Because of the approximate nature of the statistical model and the inferential procedure for the model parameters (nested sampling in our case), this may, however, not be the case in practice.

To test the "well calibratedness" of TD-CARMA's error bars, and to verify that the posterior distributions for $\Delta$ inferred by TD-CARMA on real data can be trusted, we compute the empirical coverage of $100(1-\alpha)\%$ confidence intervals for all $\alpha \in [0, 1]$ as follows. We simulate $N = 300$ doubly lensed data sets under our model (as detailed in Section 4) with true time delay $\Delta_{\mathrm{true}}$. Then, we fit TD-CARMA to each of these data sets. For a given confidence level $100(1-\alpha)\%$, we compute the $100(1-\alpha)\%$ confidence intervals derived by TD-CARMA on each simulated data set, and derive the empirical coverage as the proportion of those intervals that contain $\Delta_{\text{true}}$. We repeat this procedure for every $\alpha \in [0, 1]$, and plot the empirical coverage against the corresponding $\alpha$ (see Figure 3). Finally, we visually inspect the resulting plot, and if the empirical coverage matches the theoretical coverage (i.e., if the plotted curve is "close enough" to the 45° line representing perfect calibration), we conclude that our model produces well-calibrated error bars.

## Appendix C
## Data Set Application Results

This section reports the summary statistics for TD-DRW and TD-CARMA fitted to the doubly lensed quasar data sets of HS2209+1914 (Tables 11 and 12), SDSS J1001+5027 (Tables 13 and 14), SDSS J1206+4332 (Tables 15 and 16), SDSS J1515+1511 (Tables 17 and 18), SDSS J1455+1447 (Tables 19 and 20), and SDSS J1349+1227 (Tables 21 and 22).

**Table 11**
Summary Statistics for the Models and Posterior Modes That Do Not Detect a (QPO) Frequency in the (Posterior) PSD of the Intrinsic Light Curve Model for the HS2209+1914 Doubly Lensed Quasar

| (p, q, m) | M/n | $\ln(\mathcal{Z})$ | $\hat{\Delta}$ | SD($\hat{\Delta}$) |
|---|---|---|---|---|
| (2, 1, 1) | 1/1 | 2714.62 | −42.32 | 2.430 |
| (2, 1, 2) | 1/1 | 2723.52 | −41.82 | 2.561 |
| (2, 1, 3) | 1/1 | 2752.52 | −21.74 | 1.423 |
| (3, 1, 1) | 1/2 | 2714.37 | −43.03 | 1.952 |
| (3, 1, 2) | 3/3 | 2723.00 | −42.09 | 2.741 |
| (3, 1, 3) | 1/2 | 2751.56 | −21.73 | 1.436 |
| (3, 2, 1) | 1/1 | 2719.65 | −38.16 | 3.324 |
| (3, 2, 2) | 1/1 | 2729.66 | −37.21 | 3.466 |
| (3, 2, 3) | 1/2 | 2759.78 | −21.96 | 1.420 |
| (3, 2, 3) | 2/2 | 2759.23 | −21.94 | 1.460 |
| (4, 2, 3) | 1/3 | 2759.83 | −21.95 | 1.403 |
| (4, 3, 2) | 1/1 | 2730.02 | −38.17 | 3.198 |

**Note.** The "M/n" column denotes the number of the reported mode as given by MultiNest, as well the total number of modes in the posterior distribution.

**Table 12**
Same as Table 11 for HS2209+1914, but for Models and Posterior Modes That Detect a (QPO) Frequency, the Posterior Mean of Which is Denoted by $\hat{f}$

| (p, q, m) | M/n | $\ln(\mathcal{Z})$ | $\hat{\Delta}$ | SD($\hat{\Delta}$) | $\hat{f}$ |
|---|---|---|---|---|---|
| (3, 1, 1) | 1/2 | 2798.63 | −36.69 | 4.029 | 1.003 |
| (3, 1, 2) | 1/3 | 2799.77 | −36.78 | 0.791 | 1.003 |
| (3, 1, 2) | 2/3 | 2797.64 | −44.59 | 0.428 | 1.002 |
| (3, 1, 3) | 1/2 | 2799.68 | −32.95 | 4.672 | 1.002 |
| (4, 1, 1) | 1/2 | 2804.15 | −29.02 | 4.672 | 1.003 |
| (4, 1, 1) | 2/2 | 2804.08 | −28.93 | 5.184 | 1.003 |
| (4, 1, 2) | 1/2 | 2804.18 | −34.85 | 6.110 | 1.003 |
| (4, 1, 2) | 2/2 | 2803.72 | −35.82 | 6.795 | 1.003 |
| (4, 1, 3) | 1/3 | 2802.87 | −27.12 | 3.335 | 1.003 |
| (4, 1, 3) | 2/3 | 2801.53 | −26.50 | 2.917 | 1.003 |
| (4, 1, 3) | 3/3 | 2811.39 | −28.48 | 3.728 | 1.003 |
| (4, 2, 1) | 1/2 | 2800.93 | −25.55 | 0.431 | 1.003 |
| (4, 2, 1) | 2/2 | 2800.25 | −30.56 | 0.384 | 1.003 |
| (4, 2, 2) | 1/2 | 2803.36 | −34.64 | 6.367 | 1.003 |
| (4, 2, 2) | 2/2 | 2800.34 | −36.23 | 6.249 | 1.003 |
| (4, 2, 3) | 2/3 | 2792.02 | −25.22 | 1.993 | 1.003 |
| (4, 2, 1) | 3/3 | 2796.89 | −29.71 | 3.243 | 1.003 |
| (4, 3, 1) | 1/2 | 2802.54 | −35.48 | 2.254 | 1.003 |
| (4, 3, 1) | 2/2 | 2800.97 | −34.31 | 1.724 | 1.003 |
| (4, 3, 3) | 1/1 | 2807.15 | −28.75 | 4.521 | 1.003 |

**Note.** The posterior mean of the detected frequency is denoted by $\hat{f}$.

Table 13
Same as Table 11, but for the SDSS J1001+5027 Doubly Lensed Quasar Data Set, Models, and Posterior Modes That Do Not Detect a (QPO) Frequency

| (p, q, m) | M/n | ln($\mathcal{Z}$) | $\hat{\Delta}$ | SD($\hat{\Delta}$) |
|---|---|---|---|---|
| (2, 1, 1) | 1/1 | 2740.09 | 119.94 | 0.875 |
| (2, 1, 2) | 1/1 | 2743.91 | 120.70 | 0.799 |
| (2, 1, 3) | 1/1 | 2744.05 | 120.18 | 0.749 |
| (3, 1, 1) | 1/2 | 2739.26 | 120.05 | 0.797 |
| (3, 1, 2) | 1/2 | 2743.17 | 120.66 | 0.753 |
| (3, 1, 3) | 1/2 | 2743.06 | 120.23 | 0.686 |
| (3, 2, 1) | 1/2 | 2754.78 | 120.16 | 1.076 |
| (3, 2, 1) | 2/2 | 2753.80 | 119.87 | 1.117 |
| (3, 2, 2) | 1/1 | 2759.44 | 120.88 | 1.013 |
| (3, 2, 3) | 1/2 | 2759.48 | 120.16 | 1.003 |
| (3, 2, 3) | 2/2 | 2759.48 | 120.16 | 1.040 |
| (4, 2, 1) | 1/1 | 2753.89 | 119.94 | 1.025 |
| (4, 2, 2) | 1/1 | 2758.21 | 120.85 | 0.939 |
| (4, 2, 3) | 1/1 | 2758.35 | 120.21 | 0.970 |
| (4, 3, 1) | 1/1 | 2756.21 | 119.85 | 1.041 |
| (4, 3, 2) | 1/1 | 2761.25 | 120.93 | 1.015 |
| (4, 3, 3) | 1/1 | 2761.12 | 120.21 | 1.018 |

Table 14
Same as Table 11, but for the SDSS J1001+5027 Doubly Lensed Quasar Data Set, Models, and Posterior Modes That Detect a (QPO) Frequency

| (p, q, m) | M/n | ln($\mathcal{Z}$) | $\hat{\Delta}$ | SD($\hat{\Delta}$) | $\hat{f}$ |
|---|---|---|---|---|---|
| (3, 1, 1) | 2/2 | 2750.12 | 120.36 | 0.134 | 2.00 |
| (3, 1, 2) | 2/2 | 2743.17 | 120.42 | 0.217 | 2.00 |
| (3, 1, 3) | 1/2 | 2751.63 | 120.36 | 0.224 | 2.00 |
| (4, 1, 1) | 1/2 | 2748.84 | 120.37 | 0.112 | 2.00 |
| (4, 1, 1) | 2/2 | 2748.96 | 120.36 | 0.114 | 2.00 |
| (4, 1, 2) | 1/3 | 2750.95 | 120.38 | 0.018 | 2.00 |
| (4, 1, 2) | 2/3 | 2747.68 | 120.88 | 0.010 | 2.00 |
| (4, 1, 2) | 3/3 | 2750.97 | 120.41 | 0.177 | 2.00 |
| (4, 1, 3) | 1/2 | 2751.07 | 120.36 | 0.187 | 2.00 |
| (4, 1, 3) | 2/2 | 2750.92 | 120.36 | 0.170 | 2.00 |

Table 15
Same as Table 11, but for the SDSS J1206+4332 Doubly Lensed Quasar Data Set, Models, and Posterior Modes That Do Not Detect a (QPO) Frequency

| (p, q, m) | M/n | ln($\mathcal{Z}$) | $\hat{\Delta}$ | SD($\hat{\Delta}$) |
|---|---|---|---|---|
| (2, 1, 1) | 1/1 | 1562.55 | 112.50 | 1.454 |
| (2, 1, 2) | 1/1 | 1616.64 | 108.33 | 2.240 |
| (2, 1, 3) | 1/1 | 1624.65 | 112.78 | 1.693 |
| (3, 1, 1) | 2/2 | 1560.95 | 112.72 | 1.379 |
| (3, 1, 2) | 2/2 | 1613.46 | 108.44 | 2.259 |
| (3, 1, 3) | 2/2 | 1623.66 | 112.77 | 1.670 |
| (3, 2, 1) | 1/3 | 1573.20 | 112.79 | 1.288 |
| (3, 2, 2) | 1/2 | 1628.09 | 109.27 | 1.285 |
| (3, 2, 2) | 2/2 | 1627.21 | 109.27 | 1.382 |
| (3, 2, 3) | 1/2 | 1638.76 | 111.50 | 1.389 |
| (3, 2, 3) | 2/2 | 1637.71 | 111.50 | 1.469 |
| (4, 2, 1) | 1/3 | 1572.10 | 112.79 | 1.400 |

**Table 16**
Same as Table 11, but for the SDSS J1206+4332 Doubly Lensed Quasar Data Set, Models, and Posterior Modes That Detect a (QPO) Frequency, the Posterior Mean of Which Is Denoted by $\hat{f}$

| (p, q, m) | M/n | ln($\mathcal{Z}$) | $\hat{\Delta}$ | SD($\hat{\Delta}$) | $\hat{f}$ |
|---|---|---|---|---|---|
| (3, 1, 1) | 1/2 | 1641.00 | 107.14 | 2.332 | 1.000 |
| (3, 1, 2) | 1/2 | 1666.11 | 110.65 | 0.970 | 1.000 |
| (3, 1, 3) | 1/2 | 1667.05 | 111.07 | 0.974 | 1.000 |
| (3, 2, 1) | 2/3 | 1629.86 | 109.71 | 1.255 | 1.000 |
| (3, 2, 1) | 3/3 | 1572.42 | 112.79 | 1.628 | 1.000 |
| (4, 1, 1) | 1/2 | 1641.07 | 107.63 | 2.459 | 1.000 |
| (4, 1, 1) | 2/2 | 1641.08 | 107.66 | 2.428 | 1.000 |
| (4, 1, 2) | 1/2 | 1671.95 | 110.64 | 0.934 | 1.000 |
| (4, 1, 2) | 2/2 | 1665.42 | 110.60 | 1.039 | 1.000 |
| (4, 1, 3) | 1/2 | 1666.43 | 111.11 | 0.965 | 1.000 |
| (4, 1, 3) | 2/2 | 1672.39 | 111.23 | 0.875 | 1.000 |
| (4, 2, 1) | 2/3 | 1635.27 | 109.18 | 0.0207 | 1.000 |
| (4, 2, 1) | 3/3 | 1634.54 | 108.20 | 0.0196 | 1.000 |
| (4, 2, 2) | 1/3 | 1669.24 | 111.06 | 0.0230 | 1.000 |
| (4, 2, 2) | 2/3 | 1668.09 | 110.06 | 0.0201 | 1.000 |
| (4, 2, 2) | 3/3 | 1663.48 | 110.72 | 0.871 | 1.000 |
| (4, 2, 3) | 1/2 | 1666.03 | 111.15 | 0.969 | 1.000 |
| (4, 2, 3) | 2/2 | 1674.30 | 111.25 | 0.844 | 1.000 |
| (4, 3, 1) | 1/2 | 1643.52 | 108.43 | 1.766 | 1.000 |
| (4, 3, 1) | 2/2 | 1637.00 | 108.10 | 1.346 | 1.000 |
| (4, 3, 2) | 1/2 | 1649.08 | 106.11 | 0.865 | 1.000 |
| (4, 3, 2) | 2/2 | 1671.06 | 110.66 | 0.752 | 1.000 |
| (4, 3, 3) | 1/2 | 1673.05 | 111.25 | 0.713 | 1.000 |
| (4, 3, 3) | 2/2 | 1671.18 | 111.22 | 0.722 | 1.000 |

**Table 17**
Same as Table 11, but for the SDSS J1515+1511 Doubly Lensed Quasar Data Set, Models, and Posterior Modes That Do Not Detect a (QPO) Frequency

| (p, q, m) | M/n | ln($\mathcal{Z}$) | $\hat{\Delta}$ | SD($\hat{\Delta}$) |
|---|---|---|---|---|
| (2, 1, 1) | 1/1 | 1262.81 | 211.22 | 1.953 |
| (2, 1, 2) | 1/2 | 983.69 | −244.07 | 0.559 |
| (2, 1, 2) | 2/2 | 1257.34 | 211.23 | 2.184 |
| (2, 1, 3) | 1/2 | 1249.84 | 211.21 | 2.108 |
| (2, 1, 3) | 2/2 | 1004.07 | −292.94 | 2.360 |
| (3, 1, 1) | 1/1 | 1263.43 | 211.37 | 2.013 |
| (3, 1, 2) | 1/5 | 985.79 | −243.97 | 0.370 |
| (3, 1, 2) | 4/5 | 1252.47 | 210.87 | 1.591 |
| (3, 1, 2) | 5/5 | 1257.45 | 211.32 | 2.165 |
| (3, 1, 3) | 1/3 | 1004.57 | −292.85 | 2.234 |
| (3, 1, 3) | 3/3 | 1251.68 | 211.40 | 2.165 |
| (3, 2, 1) | 1/1 | 1274.76 | 210.66 | 2.131 |
| (3, 2, 2) | 1/2 | 1268.86 | 210.61 | 2.452 |
| (3, 2, 2) | 1/2 | 1021.76 | −251.84 | 2.507 |
| (3, 2, 3) | 1/2 | 1031.68 | −250.56 | 2.487 |
| (3, 2, 3) | 2/2 | 1263.12 | 210.72 | 2.476 |
| (4, 2, 1) | 1/1 | 1275.44 | 210.80 | 2.179 |
| (4, 2, 2) | 1/2 | 1023.55 | −251.82 | 2.210 |
| (4, 2, 2) | 2/2 | 1269.33 | 210.69 | 2.432 |
| (4, 2, 3) | 1/2 | 1263.33 | 210.80 | 2.492 |
| (4, 2, 3) | 2/2 | 1032.84 | −250.28 | 2.226 |
| (4, 3, 2) | 1/2 | 1270.09 | 210.60 | 2.393 |
| (4, 3, 2) | 2/2 | 1021.88 | −251.98 | 2.526 |
| (4, 3, 3) | 1/2 | 1032.56 | −250.79 | 2.465 |
| (4, 3, 3) | 2/2 | 1264.36 | 210.67 | 2.437 |

**Note.** The "M/n" column denotes the number of the reported mode as given by MultiNest, as well the total number of modes in the posterior distribution.

**Table 18**
Same as Table 11, but for the SDSS J1515+1511 Doubly Lensed Quasar Data Set, Models, and Posterior Modes That Detect a (QPO) Frequency, the Posterior mean of Which Is Denoted by $\hat{f}$

| (p, q, m) | M/n | ln($\mathcal{Z}$) | $\hat{\Delta}$ | SD($\hat{\Delta}$) | $\hat{f}$ |
|---|---|---|---|---|---|
| (3, 1, 2) | 2/5 | 1258.89 | 211.32 | 2.017 | 0.672 |
| (3, 1, 2) | 3/5 | 1262.72 | 211.46 | 2.192 | 2.002 |
| (3, 1, 3) | 2/3 | 1257.18 | 211.43 | 2.236 | 1.979 |
| (4, 1, 1) | 1/2 | 1273.04 | 210.71 | 1.636 | 1.994 |
| (4, 1, 1) | 2/2 | 1273.64 | 210.74 | 1.696 | 1.996 |
| (4, 1, 2) | 1/4 | 1159.61 | −254.00 | 2.951 | 1.002 |
| (4, 1, 2) | 2/4 | 1155.10 | −258.14 | 1.153 | 1.003 |
| (4, 1, 2) | 3/4 | 1266.85 | 210.90 | 2.287 | 1.005 |
| (4, 1, 2) | 4/4 | 1268.53 | 210.84 | 1.972 | 1.977 |
| (4, 1, 3) | 1/4 | 1185.33 | −210.86 | 16.421 | 1.002 |
| (4, 1, 3) | 2/4 | 1184.57 | −217.25 | 24.146 | 1.002 |
| (4, 1, 3) | 3/4 | 1262.77 | 210.93 | 2.061 | 1.794 |
| (4, 1, 3) | 4/4 | 1262.84 | 210.98 | 2.101 | 1.861 |

**Note.** The "M/n" column denotes the number of the reported mode as given by MultiNest, as well the total number of modes in the posterior distribution.

**Table 19**
Same as Table 11, but for the SDSS J1455+1447 Doubly Lensed Quasar Data Set, Models, and Posterior Modes That Do Not Detect a (QPO) Frequency

| (p, q, m) | M/n | ln($\mathcal{Z}$) | $\hat{\Delta}$ | SD($\hat{\Delta}$) |
|---|---|---|---|---|
| (2, 1, 1) | 1/1 | 552.04 | 45.37 | 1.874 |
| (2, 1, 2) | 1/1 | 550.46 | 45.21 | 1.915 |
| (2, 1, 3) | 1/2 | 543.10 | 45.21 | 1.942 |
| (2, 1, 3) | 2/2 | 516.24 | −146.13 | 5.964 |
| (3, 1, 1) | 2/2 | 552.44 | 45.58 | 1.878 |
| (3, 1, 2) | 3/3 | 550.54 | 45.19 | 1.877 |
| (3, 1, 3) | 2/4 | 513.82 | 173.21 | 3.473 |
| (3, 1, 3) | 3/4 | 516.60 | −145.73 | 5.159 |
| (3, 2, 1) | 1/1 | 552.72 | 45.36 | 1.927 |
| (3, 2, 2) | 1/1 | 550.99 | 45.28 | 1.917 |
| (3, 2, 3) | 1/3 | 516.36 | −145.66 | 2.903 |
| (3, 2, 3) | 3/3 | 543.70 | 45.22 | 1.970 |
| (4, 3, 3) | 1/1 | 544.43 | 45.28 | 1.981 |

**Note.** The "M/n" column denotes the number of the reported mode as given by MultiNest, as well the total number of modes in the posterior distribution.

**Table 20**
Same as Table 11, but for the SDSS J1455+1447 Doubly Lensed Quasar Data Set, Models, and Posterior Modes That Detect a (QPO) Frequency, the Posterior Mean of Which Is Denoted by $\hat{f}$

| (p, q, m) | M/n | ln($\mathcal{Z}$) | $\hat{\Delta}$ | SD($\hat{\Delta}$) | $\hat{f}$ |
|---|---|---|---|---|---|
| (3, 1, 1) | 1/2 | 551.13 | 45.49 | 1.763 | 2.005 |
| (3, 1, 2) | 1/3 | 547.88 | 44.94 | 1.055 | 0.485 |
| (3, 1, 2) | 2/3 | 549.17 | 45.57 | 0.945 | 2.569 |
| (3, 1, 3) | 1/4 | 514.36 | 173.39 | 3.653 | 1.388 |
| (3, 1, 3) | 4/4 | 516.52 | −145.15 | 4.635 | 1.440 |
| (3, 2, 3) | 2/3 | 517.43 | −144.12 | 4.422 | 0.112 |
| (4, 1, 1) | 1/3 | 551.62 | 45.79 | 1.334 | 1.708 |

**Table 20**
(Continued)

| $(p, q, m)$ | M/n | $\ln(\mathcal{Z})$ | $\hat{\Delta}$ | SD($\hat{\Delta}$) | $\hat{f}$ |
|---|---|---|---|---|---|
| (4, 1, 1) | 2/3 | 551.49 | 45.79 | 1.325 | 1.578 |
| (4, 1, 1) | 3/3 | 510.95 | −136.32 | 10.961 | 1.984 |
| (4, 1, 2) | 1/5 | 519.55 | 34.71 | 1.034 | 1.024 |
| (4, 1, 2) | 2/5 | 523.01 | 45.01 | 1.480 | 0.990 |
| (4, 1, 2) | 3/5 | 511.23 | −94.78 | 11.932 | 1.001 |
| (4, 1, 2) | 4/5 | 549.05 | 45.08 | 1.013 | 2.176 |
| (4, 1, 2) | 5/5 | 549.10 | 45.14 | 1.138 | 1.800 |
| (4, 1, 3) | 1/4 | 517.62 | −146.65 | 4.359 | 0.225 |
| (4, 1, 3) | 2/4 | 518.42 | −147.05 | 4.290 | 0.149 |
| (4, 1, 3) | 3/4 | 516.11 | 172.09 | 3.744 | 0.123 |
| (4, 1, 3) | 4/4 | 516.22 | 172.01 | 3.812 | 0.130 |
| (4, 2, 1) | 1/2 | 552.74 | 45.94 | 0.844 | 1.189 |
| (4, 2, 1) | 2/2 | 552.70 | 45.74 | 0.869 | 1.141 |
| (4, 2, 2) | 1/2 | 550.59 | 45.31 | 0.966 | 1.332 |
| (4, 2, 2) | 2/2 | 550.73 | 45.30 | 0.868 | 1.247 |
| (4, 2, 3) | 1/4 | 519.05 | −142.77 | 3.730 | 0.111 |
| (4, 2, 3) | 2/4 | 519.35 | −144.63 | 4.535 | 0.113 |
| (4, 2, 3) | 3/4 | 517.11 | 172.38 | 3.883 | 0.111 |
| (4, 2, 3) | 4/4 | 517.01 | 172.15 | 3.701 | 0.114 |
| (4, 3, 1) | 1/2 | 552.40 | 46.51 | 0.733 | 0.115 |
| (4, 3, 1) | 1/2 | 552.74 | 46.54 | 0.703 | 0.116 |
| (4, 3, 2) | 1/2 | 550.81 | 46.14 | 0.956 | 0.134 |
| (4, 3, 2) | 2/2 | 550.88 | 46.06 | 1.120 | 0.139 |

**Note.** The "M/n" column denotes the number of the reported mode as given by MultiNest, as well the total number of modes in the posterior distribution.

**Table 21**
Same as Table 11, but for the SDSS J1349+1227 Doubly Lensed Quasar Data Set, Models, and Posterior Modes That Do Not Detect a (QPO) Frequency

| $(p, q, m)$ | M/n | $\ln(\mathcal{Z})$ | $\hat{\Delta}$ | SD($\hat{\Delta}$) |
|---|---|---|---|---|
| (2, 1, 1) | 1/2 | 697.27 | 431.17 | 2.647 |
| (2, 1, 1) | 2/2 | 663.17 | −188.23 | 2.704 |
| (2, 1, 2) | 1/2 | 691.76 | 431.13 | 2.660 |
| (2, 1, 2) | 2/2 | 657.66 | −185.51 | 2.852 |
| (2, 1, 3) | 1/2 | 689.77 | 431.31 | 2.399 |
| (2, 1, 3) | 2/2 | 654.09 | −185.82 | 3.153 |
| (3, 1, 1) | 1/3 | 663.57 | −187.96 | 2.576 |
| (3, 1, 1) | 2/3 | 698.77 | 432.05 | 1.950 |
| (3, 1, 2) | 1/5 | 659.95 | −185.86 | 0.855 |
| (3, 1, 2) | 4/5 | 692.64 | 431.86 | 2.193 |
| (3, 1, 3) | 2/4 | 690.45 | 432.02 | 1.154 |
| (3, 1, 3) | 4/4 | 654.21 | −185.99 | 0.768 |
| (3, 2, 1) | 1/1 | 698.58 | 431.09 | 2.698 |
| (3, 2, 2) | 1/1 | 692.84 | 430.82 | 2.873 |
| (3, 2, 3) | 1/1 | 691.13 | 431.11 | 2.483 |
| (4, 1, 2) | 3/3 | 693.32 | 432.12 | 2.040 |
| (4, 1, 3) | 1/3 | 690.75 | 431.82 | 2.059 |
| (4, 2, 1) | 1/1 | 698.84 | 431.65 | 2.339 |
| (4, 2, 2) | 1/1 | 693.52 | 431.33 | 2.570 |
| (4, 2, 3) | 1/1 | 691.53 | 431.34 | 2.403 |
| (4, 3, 1) | 1/1 | 698.72 | 431.42 | 2.506 |
| (4, 3, 2) | 1/1 | 693.22 | 431.15 | 2.676 |
| (4, 3, 3) | 1/1 | 691.52 | 431.32 | 2.307 |

**Note.** The "M/n" column denotes the number of the reported mode as given by MultiNest, as well the total number of modes in the posterior distribution.

**Table 22**
Same as Table 11, but for the SDSS J1349+1227 Doubly Lensed Quasar Data Set, Models, and Posterior Modes That Detect a (QPO) Frequency, the Posterior Mean of Which Is Denoted by $\hat{f}$

| $(p, q, m)$ | M/n | $\ln(\mathcal{Z})$ | $\hat{\Delta}$ | SD($\hat{\Delta}$) | $\hat{f}$ |
|---|---|---|---|---|---|
| (3, 1, 1) | 3/3 | 698.23 | 432.31 | 1.783 | 0.857 |
| (3, 1, 2) | 2/5 | 658.59 | −186.37 | 0.403 | 0.112 |
| (3, 1, 2) | 2/5 | 662.03 | −181.27 | 6.219 | 1.007 |
| (3, 1, 2) | 5/5 | 692.26 | 432.06 | 2.144 | 0.986 |
| (3, 1, 3) | 1/4 | 690.09 | 431.88 | 1.453 | 1.429 |
| (3, 1, 3) | 3/4 | 655.05 | −185.98 | 1.907 | 2.507 |
| (4, 1, 1) | 1/4 | 698.35 | 432.75 | 0.187 | 0.531 |
| (4, 1, 1) | 2/4 | 698.30 | 432.75 | 0.261 | 0.550 |
| (4, 1, 1) | 3/4 | 663.79 | −188.88 | 2.066 | 2.036 |
| (4, 1, 1) | 4/4 | 664.05 | −182.07 | 7.044 | 1.074 |
| (4, 1, 2) | 1/3 | 692.63 | 432.78 | 0.361 | 0.561 |
| (4, 1, 2) | 2/3 | 692.71 | 432.76 | 0.375 | 0.561 |
| (4, 1, 3) | 2/3 | 664.59 | −236.90 | 22.062 | 1.003 |
| (4, 1, 3) | 3/3 | 661.61 | −232.48 | 43.940 | 1.004 |

**Note.** The "M/n" column denotes the number of the reported mode as given by MultiNest, as well the total number of modes in the posterior distribution.

### ORCID iDs

Antoine D. Meyer https://orcid.org/0000-0002-9935-6829
David A. van Dyk https://orcid.org/0000-0002-0816-331X
Hyungsuk Tak https://orcid.org/0000-0003-0334-8742
Aneta Siemiginowska https://orcid.org/0000-0002-0905-7375